\documentclass{jfm}
\usepackage{graphicx}
\usepackage{epstopdf,epsfig}
\usepackage[colorlinks=true,allcolors=blue]{hyperref}
\usepackage{upgreek} 
\usepackage[export]{adjustbox} 
\usepackage{changes}
\makeatletter
\@namedef{Changes@AuthorColor}{red}
\colorlet{Changes@Color}{red}
\makeatother

\newtheorem{hypothesis}{Hypothesis}
\newtheorem{definition}{Definition}

\usepackage{amsmath,mathtools}
\usepackage{amssymb}
\newcommand{\RR}{\mathit{Re}}
\newcommand{\We}{\mathit{We}}
\newcommand{\diff}{\mathrm{d}}
\newcommand{\f}{\frac}
\newcommand{\bs}{\boldsymbol}

\shorttitle{Locality in the turbulent bubble break-up cascade}
\shortauthor{W. H. R. Chan, P. L. Johnson and P. Moin}

\title{The turbulent bubble break-up cascade. Part 1. Theoretical developments}

\author{Wai Hong Ronald Chan\aff{1},
  Perry L. Johnson\aff{1,2}
  \and Parviz Moin\aff{1}
  \corresp{\email{moin@stanford.edu}}}

\affiliation{
\aff{1}Center for Turbulence Research (CTR), Stanford University, Stanford, CA 94305, USA
\aff{2}The Henry Samueli School of Engineering, University of California, Irvine, Irvine, CA 92697, USA
}

\begin{document}

\maketitle

\begin{abstract}
Breaking waves entrain gas beneath the surface. The wave-breaking process energizes turbulent fluctuations that break bubbles in quick succession to generate a wide range of bubble sizes. Understanding this generation mechanism paves the way towards the development of predictive models for large-scale maritime and climate simulations. \citet{Garrett1} suggested that super-Hinze-scale turbulent break-up transfers entrained gas from large to small bubble sizes in the manner of a cascade. We provide a theoretical basis for this bubble-mass cascade by appealing to how energy is transferred from large to small scales in the energy cascade central to single-phase turbulence theories. A bubble break-up cascade requires that break-up events predominantly transfer bubble mass from a certain bubble size to a slightly smaller size on average. This property is called locality. In this paper, we analytically quantify locality by extending the population balance equation in conservative form to derive the bubble-mass transfer rate from large to small sizes. Using our proposed measures of locality, we show that scalings relevant to turbulent bubbly flows, including those postulated by \citet{Garrett1} and observed in breaking-wave experiments and simulations, are consistent with a strongly local transfer rate, where the influence of non-local contributions decays in a power-law fashion. These theoretical predictions are confirmed using numerical simulations in Part 2, revealing key physical aspects of the bubble break-up cascade phenomenology. Locality supports the universality of turbulent small-bubble break-up, which simplifies the development of subgrid-scale models to predict oceanic small-bubble statistics of practical importance.
\end{abstract}

\section{Introduction}\label{sec:intro}

Turbulent bubbly flows with a wide range of bubble sizes are ubiquitous in nature and engineering, including breaking waves in oceans~\citep[e.g.,][]{Blanchard1,Medwin1,Melville1}. These bubbles contribute richly to various transport phenomena with maritime and climate implications. Experiments such as those by~\citet{Deane1}, \citet{Tavakolinejad1}, \citet{Blenkinsopp1}, and~\citet{Masnadi1} have measured the bubble size distribution in breaking waves. Their data suggests that several physical mechanisms are at play at different length and time-scales in the generation and evolution of these bubbles. These observations are supported by recent numerical simulations of breaking Stokes waves by~\citet{Wang1} and~\citet{Deike1}, as well as of shear-flow free-surface turbulence by~\citet{Yu1,Yu2}. Many of these mechanisms are not well understood to date. Among various proposed mechanisms, the fragmentation of bubbles by turbulence has garnered significant interest. Turbulent fragmentation applies to fragmenting bubbles of sizes larger than the Hinze scale, where the action of turbulent fluctuations dominates the effects of surface tension~\citep{Kolmogorov3,Hinze1}. These super-Hinze-scale bubbles have Weber numbers on the order of or larger than unity. Note that most sub-Hinze-scale bubbles with Weber numbers smaller than unity are expected to be formed by distinct fragmentation mechanisms~\citep{Deane1,Kiger1,Chan3,Chan6}. For this reason, sub-Hinze-scale bubbles are not considered in detail in this work.

\citet{Kolmogorov3} and \citet{Hinze1} suggested that turbulent eddies successively break up sufficiently large gaseous cavities into bubbles of various sizes. The average break-up frequency of bubbles of size $D$ fragmenting via this mechanism has been postulated to scale as $\varepsilon^{1/3}D^{-2/3}$, where $\varepsilon$ is the characteristic rate of turbulent kinetic energy dissipation per unit mass. The concept behind this postulate is that the break-up of a bubble is facilitated by an eddy of a comparable size in its neighbourhood~\citep{Hinze1,Chan3}. It allows the break-up frequency to be directly estimated by the inverse of the corresponding eddy turn-over time. This frequency scaling is corroborated at bubble sizes sufficiently larger than the Hinze scale by break-up frequencies for various turbulent bubbly flows in the experiments described by~\citet{MartinezBazan1} and~\citet{RodriguezRodriguez1}, and preliminarily explored in the simulations by~\citet{Chan5}. \citet{Garrett1} further proposed a quasi-steady bubble break-up cascade to explain the formation of these bubbles. Here, large volumes of gas are entrained and subsequently broken up in quick succession by turbulence, leading to an approximately steady rate of gaseous mass transfer from large to small bubble sizes. \citet{Garrett1} suggested via dimensional analysis of a system with steady entrainment that this cascade yields a quasi-stationary bubble size distribution with a $D^{-10/3}$ power-law scaling. The theoretical analysis by~\citet{Filippov1} predicts a limiting form for the size distribution assuming a Markovian (memoryless) break-up process, which coincides with the $D^{-10/3}$ power-law scaling at intermediate bubble sizes and times when the break-up frequency above is assumed. A similar scaling was observed in ensemble-averaged size distributions from breaking waves at bubble sizes sufficiently larger than the Hinze scale. These include the measured distributions of~\citet{Loewen2},~\citet{Deane1},~\citet{Rojas1},~\citet{Blenkinsopp1}, and~\citet{Na1} [see also figure 1 of \citet{Deike1}], as well as the computed distributions of~\citet{Deike1} and~\citet{Chan5,Chan3}. The bubble break-up cascade is strictly only present in flows with infinite integral-scale Weber numbers where the Hinze scale is zero. However, these experimental and numerical observations suggest that the cascade hypothesis may be extended with reasonable accuracy to practical turbulent bubbly flows with sufficiently large integral-scale Weber numbers where the Hinze scale is finite but still much smaller than the integral length scale. Note, then, that the smallest fragmenting bubbles in the break-up cascade, and all subsequent references to ``small bubbles" in this work, should have sizes around or slightly larger than the Hinze scale. Note, also, that the aforediscussed scalings for the break-up frequency and the size distribution were formally derived for a statistically stationary and homogeneous system, where all statistics are invariant in space and time. However, one may assume in a system with a large separation of scales that the large-scale dynamics do not significantly influence the small- and intermediate-scale dynamics. These scalings would then also hold in small, localized regions across various turbulent bubbly flows.

The proposed and observed $D^{-2/3}$ power-law scaling for the bubble break-up frequency has traditionally been considered separately from the proposed and observed $D^{-10/3}$ power-law scaling for the bubble size distribution. This is in spite of the fact that both scaling laws were derived on the basis of related assumptions~\citep[see also][]{Chan7,Qi1}. As alluded to earlier, each of these scalings was obtained via dimensional analysis. Thus, on its own, neither of these laws provides unequivocal support to the presence of a bubble break-up cascade mechanism in turbulent bubbly flows. For example, \citet{Yu1,Yu2} have proposed alternative mechanisms contributing to similar power-law scalings in the bubble size distribution, also via dimensional analysis. To demonstrate the plausibility of a cascade mechanism, one has to show that the underlying nature of the break-up dynamics is compatible with the characteristics of a cascade. An ideal bubble break-up cascade should be size local, where bubble mass is transferred on average from large to successively smaller bubble sizes. In other words, locality is achieved when this net transfer rate across a certain bubble size primarily depends on the break-up statistics of bubbles of similar sizes. Note that locality is necessary for the dynamics at sufficiently small bubble sizes to be largely independent of the dynamics at the largest bubble sizes. Independence from the large-size dynamics enables these small- and intermediate-size dynamics to be universal in small, localized regions in a variety of turbulent bubbly flows. In order for a universal bubble break-up cascade at these small and intermediate sizes to be plausible, the aforementioned power-law scalings will need to be reasonably compatible with the aforediscussed notion of locality. This compatibility has not been demonstrated to date, mostly because a suitable tool has not been employed to assess it.

Population balance equations~\citep[and others]{Smoluchowski1,Smoluchowski2,Landau1,Melzak1,Williams2,Friedlander1,Friedlander2,Filippov1,Valentas2,Valentas1} have been used to characterize bubble break-up using a model kernel that includes both the break-up frequency and the size distribution. This makes the population balance equation a good candidate tool to demonstrate the plausibility of a universal bubble-mass cascade mechanism. However, it is not traditionally presented in conservative form~\citep{MartinezBazan3,Saveliev1}, where the size distribution is weighted by the bubble volume. This obscures the links between the model kernel and the direct movement of bubble mass from one bubble size to another~\citep[e.g.,][]{Hulburt1,Randolph2}. Visualizing this movement in bubble-size space is key to understanding and quantifying locality. Note that the conservative population balance equation should strictly be presented as a function of mass, since mass is the true quantity being conserved~\citep{Carrica1,Castro1}. However, the equation is considered as a function of volume in this work. This exploits the direct geometrical relationship between volume and size, and is equivalent to taking the incompressible limit of the mass-conserving equation. In the case of an oceanic breaking wave, for example, this is likely to be appropriate in the early wave-breaking stages, since most of the entrained bubbles would then reside near the wave surface. Care has to be taken for later stages of the wave-breaking process when smaller bubbles may be swept deep below the surface and compressibility effects may become important. In the remainder of this work, incompressibility is assumed, and the terms ``mass" and ``volume" are used interchangeably. Scale-space transport has also been recently explored by~\citet{Thiesset1} for liquid jet atomization in relation to the volume fraction field. They proposed using two-point statistics instead of the size distribution to characterize scale locality.

In this work, a novel treatment of the population balance equation is used to demonstrate that the aforediscussed power-law scalings for the bubble break-up frequency and size distribution are compatible with a bubble break-up cascade mechanism for turbulent bubbly flows. The population balance equation in conservative form is used to derive the bubble-mass transfer flux, which describes the rate of transfer of gaseous mass between bubbles of different sizes within a bubble population. The break-up flux from large to small bubble sizes may be evaluated by averaging over many binary break-up events in these flows, where it is assumed that every parent bubble breaks into exactly two children bubbles in each event. This paper analytically quantifies the degree to which the break-up flux is local in bubble-size space. The presence of locality would support the plausibility of the scalings proposed by~\citet{Garrett1}, which are founded on a cascade phenomenology. Detailed simulations may also be used to measure this flux and its locality, and will be analyzed in a companion paper (Part 2).

This work constructs analogies between this picture of turbulent bubble break-up and the ideas underlying the celebrated concept of the turbulent energy cascade~\citep{Richardson1,Kolmogorov1,Onsager1}. Inspiration is drawn from the eddy-viscosity-based spectral energy transfer models of~\citet{Obukhov1} and~\citet{Heisenberg1,Heisenberg2}, as well as the quasi-local spectral energy transfer models of~\citet{Kovasznay1} and~\citet{Pao1,Pao2}. These parallels between the turbulent bubble-mass and energy cascades, in particular the universality of both processes in small, localized regions of turbulent flows, lend legitimacy to the idea of subgrid-scale modelling of bubbles in large eddy simulations (LES) of turbulent two-phase flows, which inherently involve a large separation of scales.

This paper is organized as follows. In \S~\ref{sec:cascade}, the turbulent bubble-mass cascade is introduced in a parallel fashion to the turbulent energy cascade. Since locality is argued to be crucial for the validity of a cascade phenomenology, two measures of locality are introduced in the context of bubble-mass transfer. In \S~\ref{sec:formalism}, the mathematical formalism required to quantify this locality is introduced. This includes the distribution of bubble sizes, the conservative population balance equation describing the dynamics of the bubble size distribution, and the model binary break-up kernel in the population balance equation and the corresponding bubble-mass flux in bubble-size space. The locality of this flux is analyzed in \S~\ref{sec:locality} in the context of self-similar energy and bubble-mass transfer. In particular, scalings relevant to small, localized regions of turbulent bubbly flows are used to obtain an expression for the bubble-mass flux due to turbulent break-up. The measures of locality introduced at the end of \S~\ref{sec:cascade} are then used to elucidate the strength of locality in this flux. In \S~\ref{sec:model}, more parallels are drawn between the turbulent bubble-mass and energy cascades using existing spectral energy transfer models as a guide. These parallels may be used to guide the development of a subgrid-scale model for bubbles in LES of turbulent two-phase flows. Finally, conclusions are drawn in \S~\ref{sec:conclusions}.

\section{The features of a cascade mechanism}\label{sec:cascade}

\begin{figure}
  \centerline{\includegraphics[width=0.8\linewidth]{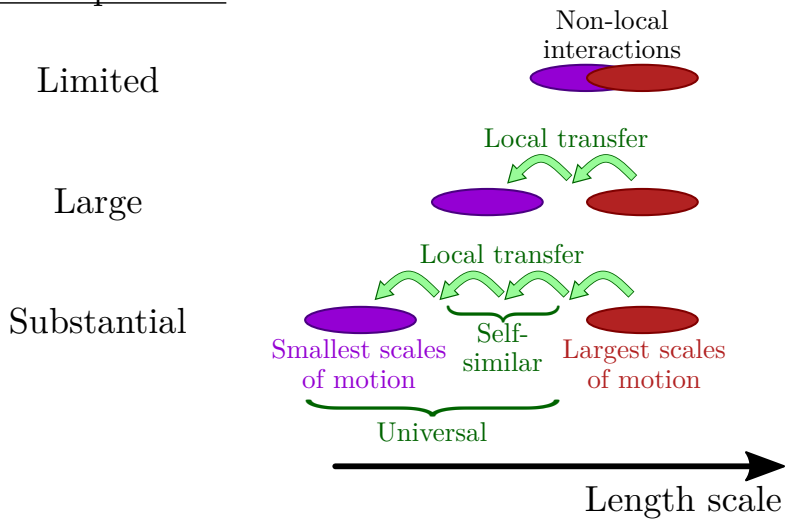}}
  \caption{Schematic illustrating the trinity of universality, locality, and self-similarity in a forward cascade. This trinity only emerges in a system with sufficient scale separation.}
\label{fig:trinity}
\end{figure}

In a forward cascade mechanism, the small- and intermediate-scale dynamics of a physical process, such as energy or bubble-mass transfer, should become independent of the large-scale flow geometry as the scale separation is increased. In other words, flow-dependent large-scale details should not directly influence the small- and intermediate-scale dynamics if there exists a clear separation of scales, and the dynamics are universal across various flows at these small and intermediate scales. This decoupling between scales suggests that the small- and intermediate-scale dynamics are scale local. When there is substantial scale separation, locality further implies that the dynamics in an intermediate subrange of scales are independent of the largest and smallest scales. Because no characteristic scale can be present in this intermediate subrange, the corresponding dynamics must be self-similar with some degree of scale invariance. This trinity of universality, locality, and self-similarity is schematically illustrated in figure~\ref{fig:trinity}. These classical ideas are reviewed for the well-established turbulent energy cascade in \S~\ref{sec:energycascade}. \citet{Garrett1} briefly alluded to a similar process for gaseous mass transfer in turbulent bubbly flows, which is examined in \S~\ref{sec:bubblecascade} with deliberate parallels to \S~\ref{sec:energycascade}. Note that these cascades hold in two scenarios: either the flow of interest and the accompanying entrainment of gas are statistically stationary, or they are quasi-steady over time-scales longer than those associated with turn-over and break-up of most of the relevant eddies and bubbles, respectively. Quasi-steadiness may be assumed in small, localized regions of turbulent flows with a sufficient separation of scales. Locality of the bubble-mass transfer in bubble-size space is vital to this cascade phenomenology. \S~\ref{sec:localityintro} discusses how locality may be quantified for the bubble-mass transfer flux, $W_b$.

\subsection{The turbulent energy cascade}\label{sec:energycascade}

The turbulent energy cascade in incompressible high-Reynolds-number single-phase flows is approximately initiated at the integral length scale $L$, i.e., the size of the largest turbulent motions, and is approximately terminated at the Kolmogorov length scale $L_\mathrm{K}$, i.e., the size of the smallest turbulent motions. Consider, at some characteristic length scale $L_n$, the characteristic inertial momentum flux $\rho_l u_{L_n}^2$, and the characteristic viscous stress $\mu_l u_{L_n} / L_n$. Here, $\rho_l$ and $\mu_l$ refer to the density and dynamic viscosity of the fluid, respectively, where the subscript $l$ assumes without loss of generality that the bulk flow involves a liquid, and $u_{L_n}$ refers to the magnitude of the characteristic velocity fluctuations associated with the length scale $L_n$. In turbulent flows, the large scales are dominated by inertial effects, while the small scales are dominated by viscous effects. The cross-over point $L_n = L_\mathrm{K}$ occurs where the characteristic inertial momentum flux approximately balances the characteristic viscous stress, such that the Reynolds number
\begin{equation}
\RR_{L_n} = \f{\rho_l u_{L_n} L_n}{\mu_l}
\label{eqn:Ren}
\end{equation}
satisfies $\RR_{L_n} = \RR_{L_\mathrm{K}} \sim O(1)$. Applying the scaling $u_{L_n} \sim \left(\varepsilon L_n\right)^{1/3}$, which holds in the inertial subrange defined by $L_\mathrm{K} \ll L_n \ll L$, and is asymptotically valid at $L_n \sim L_\mathrm{K}$, leads to the following dimensional expression for the Kolmogorov length scale 
\begin{equation}
L_\mathrm{K} \sim \left(\f{\mu_l}{\rho_l}\right)^{3/4} \varepsilon^{-1/4}.
\label{eqn:kolmodim}
\end{equation}
Note that $L_\mathrm{K}$ is a function of only $\nu_l=\mu_l/\rho_l$ and $\varepsilon$. At these small scales, the rate of energy input from the large scales $\varepsilon$ is approximately balanced by the rate of viscous dissipation $\nu_l u_{L_\mathrm{K}}^2 / L_\mathrm{K}^2$. After non-dimensionalizing $L_\mathrm{K}$ by $L$ and assuming that the energy cascade rate is dictated by the energy-containing scales $\varepsilon \sim u_L^3 / L$, one may further obtain
\begin{equation}
\f{L_\mathrm{K}}{L} \sim \RR_L^{-3/4}.
\label{eqn:kolmonondim}
\end{equation}
Taken together, these relations paint the following physical picture of the turbulent energy cascade, which was first mooted by~\citet{Richardson1} and then reiterated by~\citet{Kolmogorov1} and~\citet{Onsager1}: in a system with a sufficiently high integral-scale Reynolds number $\RR_L$, turbulent kinetic energy is cascaded from the largest to the smallest scales of turbulent motion at a rate $\varepsilon$ that is governed only by the large scales and does not vary with scale in a subrange of intermediate scales. \citet{Kolmogorov1} advanced a number of similarity hypotheses to convey these ideas for turbulent kinetic energy transfer in eddy-size space, which are recapitulated in appendix~\ref{app:energyhypo}. Note that the turbulent energy cascade is strictly valid only in the limit of zero $\nu_l$ and infinite $\RR_L$, such that $L_\mathrm{K}$ is zero. However, it may be extended with reasonable accuracy to practical turbulent flows with sufficiently large $\RR_L$, such that $L_\mathrm{K}$ is finite but still much smaller than $L$, with the understanding that the scale-invariant transfer of turbulent kinetic energy is an adequate description only in the inertial subrange $L_\mathrm{K} \ll L_n \ll L$.

For breaking waves, the magnitude of $L_\mathrm{K}$ may be estimated using the wavelength to estimate $L$, and the corresponding wave phase velocity $(gL)^{1/2}/(2\upi)^{1/2}$ to estimate $u_L$, where $g$ is the magnitude of standard gravity. For a more detailed discussion, including the potential impact of the wave slope on the estimation of the characteristic scales, see appendix B of Part 2. This yields $\RR_L^{-3/4} \sim 3 \times 10^{-5}$ for a 1-m-long wave. For the 27-cm-long waves simulated in Part 2, the corresponding dimensionless Kolmogorov length scale is $\RR_L^{-3/4} \sim 1 \times 10^{-4}$. In both cases, $L_\mathrm{K} \approx 30\text{ }\upmu\text{m}$.

\subsection{The turbulent bubble-mass cascade}\label{sec:bubblecascade}

\begin{figure}
  \centerline{\includegraphics[width=0.7\linewidth]{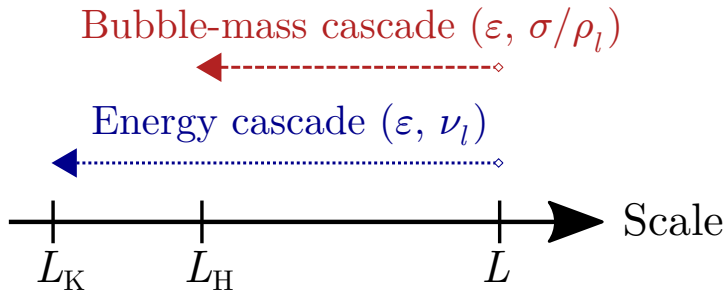}}
  \caption{Schematic illustrating the forward energy and bubble-mass cascades in turbulent bubbly flows.}
\label{fig:cascade}
\end{figure}

The turbulent bubble break-up cascade in high-Reynolds-number, high-Weber-number, incompressible, and immiscible two-phase flows is approximately initiated at $L$, i.e., the size of the largest bubbles, and is approximately terminated at the Hinze scale $L_\mathrm{H}$, i.e., the size of the smallest bubbles subject to turbulent break-up. Consider, at some characteristic length scale $L_n$, the characteristic inertial momentum flux $\rho_l u_{L_n}^2$, and the characteristic capillary pressure $\sigma/D_{L_n}$ associated with a bubble of size $D_{L_n}$ that is most relevant to the system dynamics at this length scale. Here, $\sigma$ refers to the surface tension coefficient of the gas--liquid interface. If one assumes that a bubble interacts most strongly with an eddy of the same size, then $D_{L_n} = L_n$. A physical justification for this assumption was offered by~\citet{Hinze1} and refined by~\citet{Chan3}. The cross-over point $L_n = L_\mathrm{H}$ between the large scales where inertial effects are dominant and the small scales where capillary effects are dominant occurs where the characteristic inertial momentum flux approximately balances the characteristic capillary pressure, such that the Weber number
\begin{equation}
\We_{L_n} = \f{\rho_l u_{L_n}^2 L_n}{\sigma}
\label{eqn:Wen}
\end{equation}
satisfies $\We_{L_n} = \We_{L_\mathrm{H}} \sim O(1)$. At scales larger than the Hinze scale ($L_n > L_\mathrm{H}$ and $\We_{L_n} > 1$), the dominance of inertial forces over capillary forces has been postulated to drive the fragmentation of large gaseous cavities and bubbles~\citep{Kolmogorov3,Hinze1}. This mechanism implicitly assumes that the gaseous volume fraction in the gas--liquid mixed-phase region (void fraction) is sufficiently low that coalescence between cavities and bubbles is rare. The Hinze scale is dynamically relevant only when $L_\mathrm{K} \ll L_\mathrm{H}$, so that viscous effects have a negligible influence on bubble fragmentation. The kinematic viscosity of the dispersed gaseous phase $\nu_g$ should also be less than $\nu_l$, so that the corresponding Kolmogorov length scale in the gaseous phase is less than $L_\mathrm{K}$ in the liquid~\citep{Kolmogorov3}. In addition, it is assumed that the density of the dispersed gaseous phase $\rho_g$ is smaller than $\rho_l$, so that inertial mechanisms involving the dispersed phase may be neglected. Assuming again a sufficient separation of scales in the system of interest in order for an inertial subrange to be present in the bulk turbulence, and also that the void fraction of the mixed-phase region is sufficiently low that the turbulence statistics are not significantly modified by the presence of the bubbles, the following expression for the Hinze scale can be obtained
\begin{equation}
L_\mathrm{H} \sim \left(\f{\sigma}{\rho_l}\right)^{3/5} \varepsilon^{-2/5}.
\label{eqn:hinzedim}
\end{equation}
Note that $L_\mathrm{H}$ is a function of only $\sigma/\rho_l$ and $\varepsilon$. At these small scales, the inertial momentum flux, which scales as $\rho_l \left(\varepsilon L_\mathrm{H}\right)^{2/3}$, is approximately balanced by the capillary pressure, which scales as $\sigma / L_\mathrm{H}$. After non-dimensionalizing $L_\mathrm{H}$ by $L$ and assuming again that $\varepsilon \sim u_L^3/L$, one may further obtain~[see also \citet{Shinnar2}, \citet{Narsimhan1}, \citet{Tsouris1}, \citet{Luo1}, and \citet{Apte1}]
\begin{equation}
\f{L_\mathrm{H}}{L} \sim \We_L^{-3/5}.
\label{eqn:hinzenondim}
\end{equation}
Observe the parallels between these statements and the corresponding statements in \S~\ref{sec:energycascade}, and between the relations \eqref{eqn:Ren}--\eqref{eqn:kolmonondim} and \eqref{eqn:Wen}--\eqref{eqn:hinzenondim}. One might surmise that the concept of the bubble-mass cascade transferring gaseous mass from large to successively smaller bubble sizes analogously follows the energy cascade discussed in \S~\ref{sec:energycascade}, provided the bubble-mass transfer is driven by turbulent eddies. In high-$\RR_L$ and high-$\We_L$ bubbly flows, these cascades may exist simultaneously, as illustrated in figure~\ref{fig:cascade}. A similar parallel was drawn in the context of coalescence by~\citet{Friedlander1,Friedlander2}. Like the energy flux $\varepsilon$, the bubble-mass flux $W_b$ should be governed only by the large scales and should not vary with size in a subrange of intermediate sizes. While self-similarity occurs in the inertial subrange $L_\mathrm{K} \ll L_n \ll L$ in the turbulent energy cascade, it should also be present in an analogous intermediate bubble-size subrange $L_\mathrm{H} \ll L_n \ll L$ in the turbulent bubble-mass cascade. In addition, just as the transfer of energy should be interpreted in a statistical sense through the statistics of the velocity structure functions, a probabilistic interpretation of the transfer of gaseous mass across bubble sizes is warranted. This interpretation is provided by the bubble size distribution to be introduced in \S~\ref{sec:sizedist}. Finally, in the same way that the discussion in \S~\ref{sec:energycascade} may be mapped to a set of similarity hypotheses recapitulated in appendix~\ref{app:energyhypo}, a set of similarity hypotheses for turbulent bubble-mass transfer in bubble-size space corresponding to the discussion above is proposed in appendix~\ref{app:bubblehypo}. Note that the turbulent bubble-mass cascade is strictly valid only in the limit of zero $\sigma/\rho_l$ and infinite $\We_L$, such that $L_\mathrm{H}$ is zero. However, it may be extended with reasonable accuracy to practical turbulent two-phase flows with sufficiently large $\We_L$, such that $L_\mathrm{H}$ is finite but still much smaller than $L$, with the understanding that the size-invariant transfer of bubble mass is an adequate description only in the intermediate bubble-size subrange $L_\mathrm{H} \ll L_n \ll L$, i.e., for the fragmentation of super-Hinze-scale bubbles.

Aside from the assumptions listed above, the following should also hold in the turbulent bubble-mass cascade. First, large pockets of gas $(L_n \sim L)$ need to be steadily or quasi-steadily injected into a bulk volume of liquid to facilitate the transfer of bubble mass from large to small bubble sizes. Second, buoyancy and gradual dissolution may be neglected in the bubble dynamics. Third, a mechanism for the removal of small bubbles of sizes smaller than $L_\mathrm{H}$ exists to prevent their accumulation. This physical limit holds when the time-scales of the neglected secondary effects, such as coalescence, buoyancy, gradual dissolution, and the accumulation of small bubbles, exceed the flow and entrainment time scales of interest, as alluded to by~\citet{Garrett1} as well.

For breaking waves, the magnitude of $L_\mathrm{H}$ may be estimated in a similar fashion to the estimate of $L_\mathrm{K}$ in \S~\ref{sec:energycascade}. For a 1-m-long wave, one may obtain $\We_L^{-3/5} \sim 3 \times 10^{-3}$. For the 27-cm-long waves simulated in Part 2, one may similarly obtain $\We_L^{-3/5} \sim 1 \times 10^{-2}$. In both cases, $L_\mathrm{H} \approx 3\text{ mm}$ and $L_\mathrm{H}/L_\mathrm{K} \sim \We_L^{-3/5}\RR_L^{3/4} \sim 10^2$, thus satisfying the earlier assumption $L_\mathrm{K} \ll L_\mathrm{H}$. More generally, one may write 
\begin{equation}
\f{L_\mathrm{H}}{L_\mathrm{K}} \sim \left(\f{\sigma}{\mu_l u_{L_\mathrm{K}}}\right)^{3/5}.
\end{equation}
For air--water systems, the Kolmogorov velocity scale $u_{L_\mathrm{K}}$ will need to exceed $\sigma/\mu_l \sim 10^2\text{ m/s}$ in order for $L_\mathrm{K}$ to exceed $L_\mathrm{H}$. Thus, the assumption is satisfied for most terrestrial oceanic systems where the characteristic flow speed is slower than this.

\subsection{Locality in a universal framework for turbulent bubble break-up}\label{sec:localityintro}

\begin{figure}
  \centerline{\includegraphics[width=0.6\linewidth]{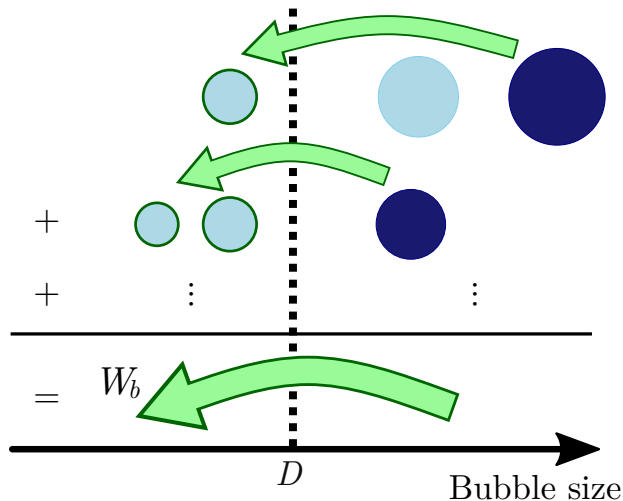}}
  \caption{Schematic illustrating the computation of the bubble break-up flux $W_b(D)$ across a particular bubble size $D$ through the appropriate averaging of gaseous mass transfers from individual events as represented by block arrows. Each row corresponds to an individual break-up event. Parent bubbles have a dark fill colour, while children bubbles have a light fill colour. Children bubbles that contribute to $W_b(D)$ are marked with a dark border. For a more comprehensive illustration, refer to figure~\ref{fig:indivevents} and the accompanying description in appendix~\ref{app:indivevents}.}
\label{fig:Wbdef}
\end{figure}

The existence of a universal cascade mechanism for bubble break-up requires the break-up process to be size local. It should be emphasized that locality of the averaged break-up dynamics\textemdash not the locality of individual break-up events\textemdash is the measure of interest since turbulent cascades should always be interpreted in a statistical manner. In order to enable this statistical interpretation, the break-up flux, $W_b$, should be derived from the averaged break-up dynamics, as illustrated in figure~\ref{fig:Wbdef}. $W_b(D)$ is the rate at which bubble mass\textemdash or, equivalently in an incompressible system, gaseous volume\textemdash is transferred from bubbles of sizes larger than $D$ to bubbles of sizes smaller than $D$, and will be introduced in more detail in \S~\ref{sec:formalism}. The link between individual break-up events and the averaged break-up dynamics is more concretely articulated through specific examples in appendix~\ref{app:indivevents}. 

Locality in $W_b$ is quantified using two complementary measures inspired by the concepts of infrared and ultraviolet locality introduced by~\citet{Lvov1} and~\citet{Eyink1} for turbulent kinetic energy transfer. First, one is interested in the degree to which incoming contributions to $W_b(D)$ from all parent bubble sizes larger than $D$ arise primarily from sizes only slightly larger than $D$. This metric is termed infrared locality, since infrared radiation has a longer wavelength than visible light. If the rate at which parent bubbles of sizes between $D_p > D$ and $D_p + \diff D_p$ transfer mass to bubbles of sizes smaller than $D$ is $I_p(D_p|D) \: \diff D_p$, then $W_b(D)$ is the integral of the incoming differential transfer rate $I_p(D_p|D)$ over all parent bubble sizes $D_p > D$. Figure \ref{fig:IRlocality}(a) illustrates this relation between $I_p$ and $W_b$. With this decomposition of $W_b$, infrared locality may then be quantified by considering how quickly the incoming differential transfer rate $I_p(D_p|D)$ from parent bubbles decays with increasing $D_p$: 

\begin{figure}
  \centerline{
(a)
\includegraphics[width=0.45\linewidth,valign=t]{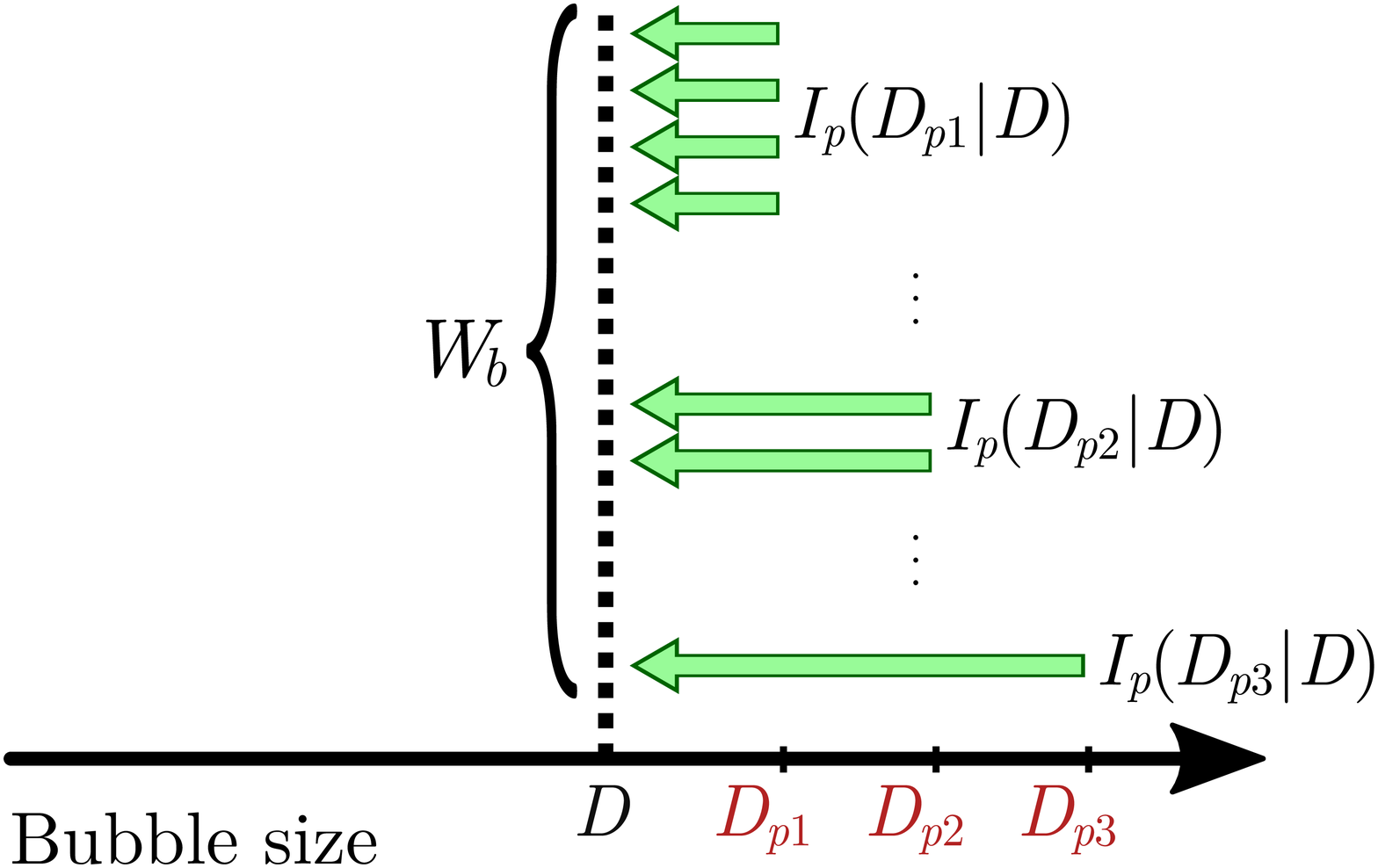}
\quad
(b)
\includegraphics[width=0.45\linewidth,valign=t]{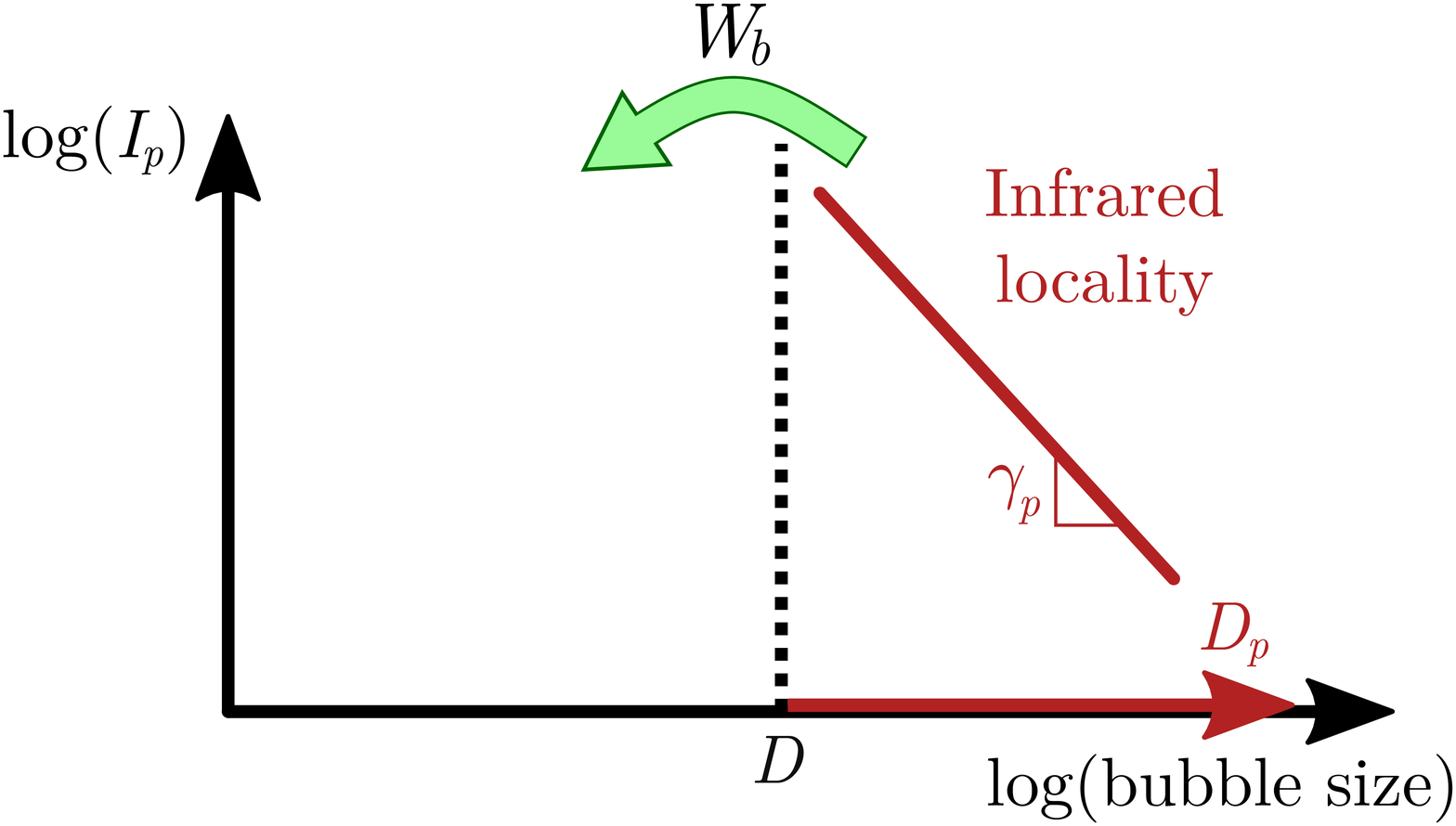}
}
  \caption{Schematics illustrating infrared locality in the break-up flux $W_b$. $W_b(D)$ may be computed by integrating the incoming differential contributions $I_p(D_p|D)$ from each parent bubble size $D_p > D$. Subfigure (a) illustrates this decomposition of $W_b$. In particular, it depicts a system where the incoming differential transfer rate $I_p(D_p|D)$ from parent bubbles varies as $I_p(D_{p1}|D) > I_p(D_{p2}|D) > I_p(D_{p3}|D)$ for $D_{p1} < D_{p2} < D_{p3}$. This variation of $I_p(D_p|D)$ with $D_p$ is graphically depicted in subfigure (b). The limiting power-law exponent $\gamma_p$ in (b) describes the behaviour $I_p(D_p\rightarrow\infty|D)$. This exponent is revisited in the relations \eqref{eqn:uniformIpinf} and \eqref{eqn:betaIpinf}.}
\label{fig:IRlocality}
\end{figure}

\begin{definition}
\normalfont
(Infrared locality) If $W_b(D)$ may be written as
\begin{equation}
W_b(D) = \int_D^\infty \diff D_p \: I_p(D_p|D),
\label{eqn:Ip}
\end{equation}
then infrared locality describes the rate at which $I_p$ decays from $D_p \sim D$ to $D_p \rightarrow \infty$. 
\label{def:IR}
\end{definition}
The variation of $I_p(D_p|D)$ with $D_p$ for an infrared local system is schematically illustrated in figure~\ref{fig:IRlocality}(b). 

Second, one is interested in the degree to which outgoing contributions to $W_b(D)$ due to all child bubble sizes smaller than $D$ are due primarily to sizes only slightly smaller than $D$. This metric is correspondingly termed ultraviolet locality. If the rate at which children bubbles of sizes between $D_c$ and $D_c + \diff D_c < D$ receive mass from bubbles of sizes larger than $D$ is $I_c(D_c|D) \: \diff D_c$, then $W_b(D)$ is the integral of the outgoing differential transfer rate $I_c(D_c|D)$ over all child bubble sizes $D_c < D$. Figure \ref{fig:UVlocality}(a) illustrates this relation between $I_c$ and $W_b$. With this decomposition of $W_b$, ultraviolet locality may then be quantified by determining how quickly the outgoing differential transfer rate $I_c(D_c|D)$ to children bubbles decays with decreasing $D_c$:

\begin{figure}
  \centerline{
(a)
\includegraphics[width=0.45\linewidth,valign=t]{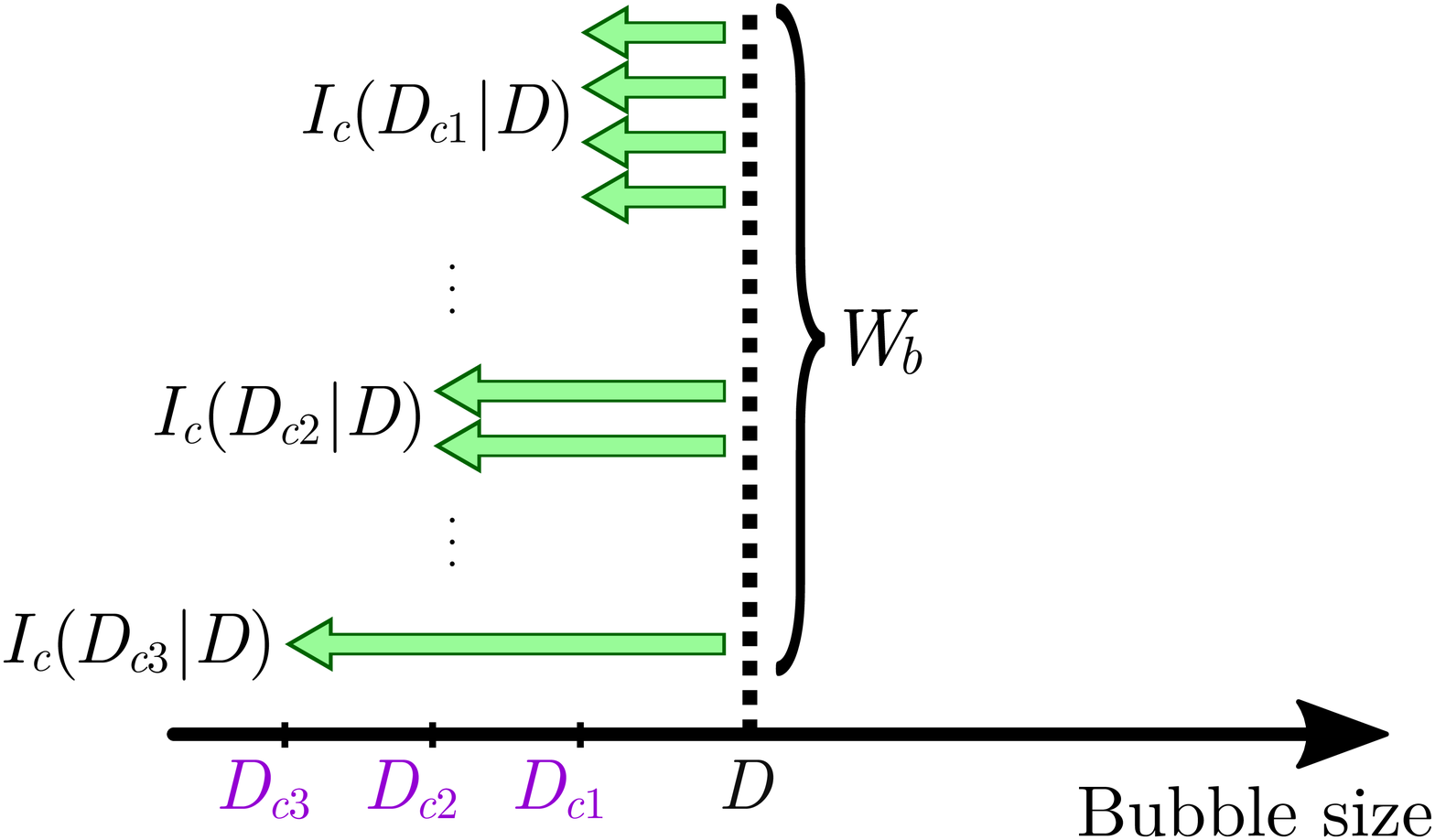}
\quad
(b)
\includegraphics[width=0.45\linewidth,valign=t]{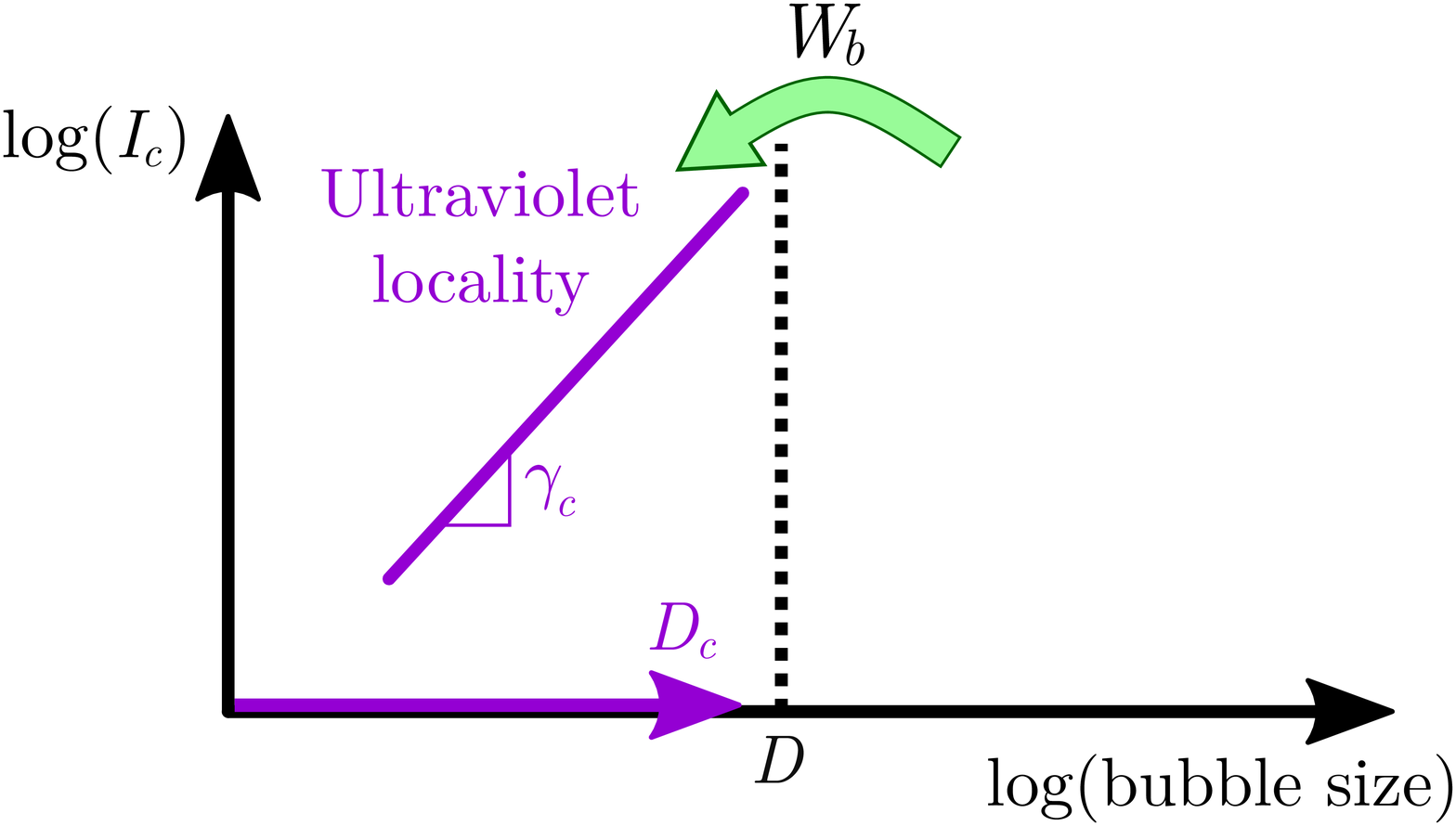}
}
  \caption{Schematics illustrating ultraviolet locality in the break-up flux $W_b$. $W_b(D)$ may be computed by integrating the outgoing differential contributions $I_c(D_c|D)$ due to each child bubble size $D_c < D$. Subfigure (a) illustrates this decomposition of $W_b$. In particular, it depicts a system where the outgoing differential transfer rate $I_c(D_c|D)$ to children bubbles varies as $I_c(D_{c1}|D) > I_c(D_{c2}|D) > I_c(D_{c3}|D)$ for $D_{c1} > D_{c2} > D_{c3}$. This variation of $I_c(D_c|D)$ with $D_c$ is graphically depicted in subfigure (b). The limiting power-law exponent $\gamma_c$ in (b) describes the behaviour $I_c(D_c\rightarrow 0|D)$. This exponent is revisited in the relations \eqref{eqn:uniformIczero} and \eqref{eqn:betaIczero}.}
\label{fig:UVlocality}
\end{figure}

\begin{definition}
\normalfont
(Ultraviolet locality) If $W_b(D)$ may be written as
\begin{equation}
W_b(D) = \int_0^D \diff D_c \: I_c(D_c|D),
\label{eqn:Ic}
\end{equation}
then ultraviolet locality describes the rate at which $I_c$ decays from $D_c \sim D$ to $D_c \rightarrow 0$. 
\label{def:UV}
\end{definition}
The variation of $I_c(D_c|D)$ with $D_c$ for an ultraviolet local system is schematically illustrated in figure~\ref{fig:UVlocality}(b). To reiterate, these decompositions of $W_b$ into $I_p$ and $I_c$ are two distinct but complementary ways of analyzing the contributions to $W_b$ from different bubble sizes. The sum of all $I_p$'s over all eligible parent bubbles $(D_p>D)$ yields $W_b(D)$, as does the sum of all $I_c$'s over all eligible children bubbles $(D_c<D)$.


\section{Mathematical formalism}\label{sec:formalism}

In this section, the bubble size distribution and its corresponding population balance equation in conservative form, together with the typical model kernel for bubble break-up, are introduced in order to derive a suitable expression for the break-up flux $W_b$, and thus the locality measures $I_p$ and $I_c$ introduced in \S~\ref{sec:localityintro}. These quantities are used in \S~\ref{sec:locality} to determine the extent of validity of the bubble-mass cascade phenomenology in \S~\ref{sec:bubblecascade}, including the proposed similarity hypotheses \ref{hyp:sphericity}--\ref{hyp:intermediatesubrange} in appendix~\ref{app:bubblehypo}.

\subsection{The bubble size distribution}\label{sec:sizedist}

At every location $\bs{x}$, for every bubble size $D$, and at some time $t$, the number density function $\mathring{f}$ for a bubble population may be constructed by adding a contribution from each bubble $i = 1, \ldots, N_b(t)$ having a centroid location $\bs{x}_i$ and an equivalent size $D_i$
\begin{equation}
\mathring{f}\left(\bs{x},D;t\right) = \sum_{i=1}^{N_b\left(t\right)} \delta\left(\bs{x}-\bs{x}_i(t)\right) \delta\left(D-D_i\left(t\right)\right),
\label{eqn:bsd_f}
\end{equation}
where $\delta$ is the Dirac delta function. Note that $\mathring{f}$ is not a probability density function since it is constructed through the accounting of bubbles in a single system snapshot. The probability distribution of bubble sizes $f$ may be obtained by ensemble averaging over statistically independent but similar realizations
\begin{equation}
f\left(\bs{x},D;t\right) = \left\langle \mathring{f}\left(\bs{x},D;t\right) \right\rangle.
\label{eqn:bsd_fbar}
\end{equation}
The probabilistic nature of this size distribution results in a break-up flux in \S~\ref{sec:kernel} that is compatible with a statistical interpretation of the break-up dynamics. Note that the dimensions of $\mathring{f}$ and $f$ are $(\text{length})^{-4}$ since the following constraints are satisfied over some sampling volume $\int_\Omega \diff \bs{x} = \mathcal{V}$ that always contains all $N_b(t)$ bubbles
\begin{equation*}
N_b\left(t\right) = \int_\Omega \diff\bs{x} \int_0^\infty \diff D \: \mathring{f}\left(\bs{x},D;t\right),\qquad 
\left\langle N_b\left(t\right) \right\rangle = \int_\Omega \diff\bs{x} \int_0^\infty \diff D \: f\left(\bs{x},D;t\right).
\end{equation*}
Here, the volume-integration $\left(\int_\Omega \diff\bs{x} \: \cdot\right)$ and ensemble-averaging $\left(\left\langle\cdot\right\rangle\right)$ operations commute only if $\Omega$ and $\mathcal{V}$ are identical over all the ensemble realizations. 

While many flows, such as breaking waves, are intrinsically statistically unsteady and inhomogeneous, smaller-scale dynamics with faster time-scales relative to larger-scale developments may evolve very similarly to statistically stationary and homogeneous flows, as suggested in hypothesis \ref{hyp:localisotropy} in appendix~\ref{app:energyhypo}. These smaller-scale dynamics occur in small, localized regions of turbulent flows with sufficient scale separation. This approximation of quasi-stationarity and quasi-homogeneity implies
\begin{equation}
f\left(\bs{x},D;t\right) \approx f(D).
\label{eqn:bsd_localisotropy}
\end{equation}
In other words, the bubble size distribution of a statistically stationary and homogeneous turbulent bubbly flow at small and intermediate bubble sizes may shed light on what might be the universal characteristics of a bubble population at small and intermediate bubble sizes in small, localized regions of turbulent bubbly flows with sufficient scale separation, and vice versa.

\subsection{The population balance equation}\label{sec:popbalance}

The population balance equation was introduced by \citet{Smoluchowski1,Smoluchowski2},~\citet{Landau1},~\citet{Melzak1},~\citet{Williams2},~\citet{Friedlander1,Friedlander2},~\citet{Filippov1},~\citet{Randolph1}, \citet{Fredrickson1}, and~\citet{Behnken1} in their respective fields. It is used here to describe the evolution of the bubble size distribution $f\left(\bs{x},D;t\right)$ in the four-dimensional phase space comprising the three spatial dimensions $\bs{x}=(x_1,x_2,x_3)$ and the bubble-size dimension $D$ as follows~\citep{Hulburt1,Randolph2}
\begin{multline}
\f{\partial \left[f\left(\bs{x},D;t\right) D^3 \right]}{\partial t} + \f{\partial \left[v_i\left(\bs{x},D;t\right) f\left(\bs{x},D;t\right) D^3 \right]}{\partial x_i} + \f{\partial \left[v_D\left(\bs{x},D;t\right) f\left(\bs{x},D;t\right) D^3 \right]}{\partial D} ={}\\
{}= H(\bs{x},D;t),
\label{eqn:pbe_fbar_orig}
\end{multline}
for some model term $H$ that includes the effects of break-up, coalescence, entrainment, and other effects. Here, $v_i$ and $v_D$ represent the velocities of the bubble-volume-weighted probability density function, $fD^3$, in phase space along the spatial and bubble-size dimensions, respectively. The $D^3$-weighting enables the equation to be written in conservative form~\citep{MartinezBazan3,Saveliev1} in the incompressible limit where mass and volume are equivalent, since bubble mass is conserved by break-up and coalescence events. Following the arguments of quasi-stationarity and quasi-homogeneity leading to \eqref{eqn:bsd_localisotropy}, one may simplify \eqref{eqn:pbe_fbar_orig} to
\begin{equation}
\underbrace{\f{\diff \left[v_D(D) f(D) D^3 \right]}{\diff D}}_{\text{local transport}} = \underbrace{H(D)}_{\substack{\text{source and sink terms,}\\\text{and non-local transport}}}.
\label{eqn:pbe_fbar}
\end{equation}
These mechanisms are schematically illustrated in figure~\ref{fig:pbe_phase}, which depicts the movement of $fD^3$ in $D$-space, and are further discussed in appendix~\ref{app:popbalancemore}. In summary, the phase-space-based form of the population balance equation, \eqref{eqn:pbe_fbar}, distinguishes the contributions of local and non-local bubble-mass transport. As explained in appendix~\ref{app:indivevents}, individual break-up (and coalescence) events are non-local in size space. However, the ensemble-averaged dynamics may be approximated as size local if they satisfy infrared and ultraviolet locality, as discussed in \S~\ref{sec:localityintro}. These concepts are appropriate particularly in the limit where the subspace of initial conditions for a bubbly system corresponding to an initial collection of large bubbles is sufficiently sampled. If a quasi-stationary limit exists for the system, then subsequent bubble break-up would lead to a continuous distribution for $f(D)$ after the transient dynamics have passed, as opposed to a discrete distribution comprising a finite number of Dirac delta functions in bubble-size space. Then, if the source and sink mechanisms are neglected, one may re-interpret the terms in \eqref{eqn:pbe_fbar} as 
\begin{equation}
\underbrace{\f{\diff \left[v_D(D) f(D) D^3 \right]}{\diff D}}_{\substack{\text{local transport approximation for}\\\text{break-up and/or coalescence}}} = \underbrace{H(D)}_{\substack{\text{error of local}\\\text{transport approximation}}}.
\label{eqn:pbe_fbar_reinterpret}
\end{equation}

\begin{figure}
  \centerline{
\includegraphics[width=0.7\linewidth]{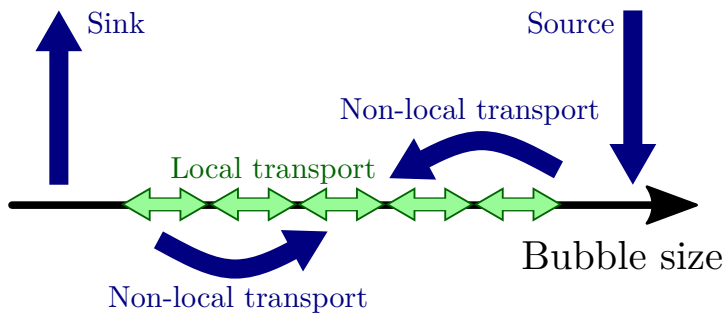}
}
  \caption{Schematic illustrating the physical significance of the terms in \eqref{eqn:pbe_fbar}. Local transport denoted by the lightly shaded block arrows corresponds to the left-hand-side term, while the remaining mechanisms denoted by the dark block arrows correspond to the right-hand-side term.}
\label{fig:pbe_phase}
\end{figure}

The population balance equation \eqref{eqn:pbe_fbar_orig} is often alternatively written, in the limit where mass-transfer processes such as dissolution that would cause individual bubble sizes to continuously increase or decrease with time may be neglected, as
\begin{multline}
\f{\partial \left[f\left(\bs{x},D;t\right) D^3 \right]}{\partial t} + \f{\partial \left[v_i\left(\bs{x},D;t\right) f\left(\bs{x},D;t\right) D^3 \right]}{\partial x_i} ={}\\
{}= T_b\left(\bs{x},D;t\right) + T_c\left(\bs{x},D;t\right) + T_s\left(\bs{x},D;t\right)
\label{eqn:pbe_fbar2_orig}
\end{multline}
for some model terms $T_b$, $T_c$, and $T_s$ corresponding to break-up, coalescence, and other sources and sinks, respectively. Once again, \eqref{eqn:pbe_fbar2_orig} may be simplified to
\begin{equation}
0 = \underbrace{T_b(D)}_{\text{break-up}} + \underbrace{T_c(D)}_{\text{coalescence}} +  \underbrace{T_s(D)}_{\substack{\text{other sources}\\\text{and sinks}}}.
\label{eqn:pbe_fbar2}
\end{equation}
The kernel-based form of the population balance equation \eqref{eqn:pbe_fbar2} isolates the contributions to bubble-mass transport from individual physical processes. Since break-up and coalescence processes do not create or destroy bubble mass, or bubble volume in the incompressible limit, $T_b(D)$ and $T_c(D)$ must individually satisfy the conservation of bubble mass; for example
\begin{equation}
\int_0^\infty \diff D \: T_b(D) = 0.\label{eqn:Tbcons}
\end{equation}
Recalling the assumptions in \S~\ref{sec:bubblecascade}, $T_c(D)$ is assumed to be negligible, while $T_s(D)$ is assumed to be active only at small $(D < L_\mathrm{H})$ and large $(D \sim L)$ bubble sizes. Thus, at intermediate bubble sizes, only $T_b(D)$ is in play. The common model kernel for $T_b(D)$ is the subject of the next subsection. At these intermediate sizes, one may compare~\eqref{eqn:pbe_fbar_reinterpret} with~\eqref{eqn:pbe_fbar2} to approximately obtain, in the limit of size-local break-up,
\begin{equation}
\underbrace{-\f{\diff \left[v_D(D) f(D) D^3 \right]}{\diff D}}_{\text{local transport approximation}} = \underbrace{T_b(D)}_{\text{break-up}}.
\label{eqn:pbe_locality}
\end{equation}
The bubble break-up process may then be modelled by an appropriate velocity in bubble-size space, $v_D(D)$, as will be further discussed in \S~\ref{sec:spectralenergy}. Quasi-stationarity and quasi-homogeneity also imply that both terms in \eqref{eqn:pbe_locality} are zero. In other words, the rate of increase of the number of bubbles of size $D$ due to the break-up of larger bubbles is dynamically balanced by the rate of decrease due to break-up into smaller bubbles. It is further shown in \S~\ref{sec:locality} that this corresponds to the self-similarity of $W_b(D)$ in the intermediate bubble-size subrange $L_\mathrm{H} \ll D \ll L$, which emerges when there is a sufficient separation of scales.

\subsection{The model binary break-up kernel and the corresponding break-up flux}\label{sec:kernel}

Assuming that all break-up events are independent of one another, i.e., that they follow a Markovian (memoryless) stochastic process where each break-up event is independent of all previous events, and that only binary break-up events occur, a model form for the break-up kernel $T_b(D)$ may be constructed as follows~\citep[e.g.,][]{Filippov1,Valentas2,Valentas1,Coulaloglou1,Ramkrishna1,MartinezBazan1,MartinezBazan2,Chan5}
\begin{equation}
T_b(D) = \int_D^\infty \diff D_p \: q_b(D|D_p) g_b(D_p) f(D_p) D^3 - g_b(D) f(D) D^3.
\label{eqn:Tb}
\end{equation}
The first term on the right-hand side is a source (birth) term due to the break-up of bubbles of sizes larger than $D$, while the second term is a sink (death) term due to the break-up of bubbles of size $D$ into smaller bubbles. The differential break-up rate $g_b (D) f (D)$ is the expected differential rate of break-up events per unit domain volume for bubbles of size $D$, which is \emph{modelled} as being proportional to the average number of bubbles of size $D$ per unit domain volume and unit size, $f(D)$. Then, $g_b(D)$ is the characteristic break-up frequency of a bubble of size $D$. Also, $q_b(D|D_p)$ is the probability that a bubble of size $D_p$ breaks into a bubble of size $D$ and another bubble of complementary volume such that the total gaseous volume remains constant through the break-up event. Several properties of $q_b(D_c|D_p)$ that will facilitate subsequent derivations are introduced in appendix~\ref{app:kernelmore}. Non-binary break-up events are addressed in appendix~\ref{app:caveats-binary}.

The corresponding break-up flux $W_b(D)$ may be interpreted in two complementary ways, recalling the concepts introduced at the end of \S~\ref{sec:localityintro}. First, it describes the net loss of mass from bubbles of sizes larger than $D$ due to the break-up process modelled by $T_b(D)$, if one decomposes $W_b$ into its incoming contributions from various parent bubbles. Second, it describes the net gain in mass in bubbles of sizes smaller than $D$, if one decomposes $W_b$ into its outgoing contributions to various children bubbles. From \eqref{eqn:Tbcons}, it is evident that these two quantities are equal in magnitude, leading to the following equivalent definitions for the break-up flux
\begin{equation}
W_b(D) = \int_0^D \diff D_c \: T_b(D_c) = -\int_D^\infty \diff D_p \: T_b(D_p).
\end{equation}
Note that this implies in turn that
\begin{equation}
\f{\diff W_b(D)}{\diff D} = T_b(D).
\label{eqn:WbTb}
\end{equation}
Observe the parallels between \eqref{eqn:pbe_locality} and \eqref{eqn:WbTb}, which will be addressed in \S~\ref{sec:spectralenergy}. 

One may show that $W_b$ satisfies
\begin{equation}
W_b(D) = \int_0^D \diff D_c \: D_c^3 \int_D^\infty \diff D_p \: q_b \left(D_c|D_p\right) g_b(D_p) f(D_p).
\label{eqn:Wb}
\end{equation}
A detailed derivation is provided in appendix~\ref{app:kernelmore}. Note that the dimensions of $W_b$ are $(\text{time})^{-1}$. Note, also, that $W_b(D)$ has been expressed in terms of integrals with limits involving $D$, similar to the expressions \eqref{eqn:Ip} and \eqref{eqn:Ic}. One may then directly infer that
\begin{gather}
I_p(D_p|D) = \int_0^D \diff D_c \: D_c^3 q_b \left(D_c|D_p\right) g_b(D_p) f(D_p),\label{eqn:WbIRI}\\
I_c(D_c|D) = \int_D^\infty \diff D_p \: D_c^3 q_b \left(D_c|D_p\right) g_b(D_p) f(D_p).\label{eqn:WbUVI}
\end{gather}
The analysis of the constituent terms in these quantities is the subject of the next section.

It is emphasized again that the break-up flux $W_b$ averages the transfer of bubble mass over many break-up events through the ensemble-averaging operation discussed in \S~\ref{sec:sizedist}. Assuming each event occurs independently, \eqref{eqn:Wb} may be interpreted as the summation of the bubble-mass (or gaseous volume) transfer $q_b(D_c|D_p) D_c^3$, multiplied by the average differential break-up rate $g_b(D_p)f(D_p)$, over all relevant parent and child bubble sizes. This is reiterated using concrete examples in appendix~\ref{app:indivevents}.


\section{Locality in bubble-mass transfer across bubble-size space}\label{sec:locality}

The presence of locality, and hence cascade-like behaviour, in the break-up flux $W_b$ driven by turbulence may be analyzed using scalings for the constituent model terms $g_b (D_p) f(D_p)$ and $q_b(D_c|D_p)$ suitable for turbulent bubble fragmentation. Consider, first, the variation of the differential break-up rate $g_b(D_p) f(D_p)$ with the parent bubble size $D_p$. As discussed in \S~\ref{sec:intro} and as presented by \citet{Chan5,Chan3}, the bubble size distribution has been theoretically, experimentally, and numerically demonstrated to scale as $D_p^{-10/3}$ for parent bubbles of a set of intermediate sizes $L_\mathrm{H} \ll D_p \ll L$ where fragmentation occurs due to turbulence in the carrier phase. The $D_p^{-10/3}$ power-law scaling for the size distribution is revisited in Part 2 in relation to a set of numerical simulations of breaking waves to be discussed. Assuming
\begin{equation}
f(D_p) \sim D_p^{-10/3}
\end{equation}
in this range of bubble sizes, it remains to examine the scaling of the characteristic break-up frequency $g_b$ with $D_p$. This may be estimated by recalling from \S~\ref{sec:energycascade} that at some length scale $D_p$ in the inertial subrange $L_\mathrm{K} \ll D_p \ll L$, turbulent velocity fluctuations scale as $u_{D_p} \sim D_p^{1/3}$. The characteristic break-up frequency of super-Hinze-scale bubbles of size $D_p$ may then be estimated as the inverse of the corresponding eddy turn-over time
\begin{equation}
g_b(D_p) \sim u_{D_p} / D_p \sim D_p^{-2/3}.
\label{eqn:gb_theory}
\end{equation}
This yields the following scaling for the differential break-up rate $g_b f$ in the intermediate size subrange $L_\mathrm{H} \ll D_p \ll L$ \citep{Filippov1,Chan7,Qi1}
\begin{equation}
g_b(D_p) f(D_p) \sim D_p^{-4}.
\end{equation}
The frequency scaling $g_b \sim D_p^{-2/3}$ has been suggested in other studies, including the break-up models of~\citet{Coulaloglou1},~\citet{Lee1,Lee2}, and~\citet{MartinezBazan1}. In addition,~\citet{MartinezBazan3} and~\citet{Qi1} demonstrated that several other models in the literature that may not at first seem to have a $D_p^{-2/3}$ scaling do in fact predict a very similar scaling at sufficiently large $D_p$. As mentioned in \S~\ref{sec:intro}, this frequency scaling was also observed in experiments discussed by~\citet{MartinezBazan1} and~\citet{RodriguezRodriguez1}, and is also consistent with the breaking-wave simulations to be discussed in Part 2. It is emphasized here that $g_b \sim D_p^{-2/3}$ is an appropriate scaling only for bubbles in the intermediate size subrange $L_\mathrm{H} \ll D_p \ll L$ where the action of turbulent velocity fluctuations dominates the effects of surface tension for the purposes of fragmentation. Thus, bubbles of sizes very close to the Hinze scale may have breakup frequencies that diverge from this idealized scaling as capillary effects enter the picture. Note, also, that the ratio $u_{D_p}/D_p$ in \eqref{eqn:gb_theory} may still be used to estimate $g_b(D_p)$ for turbulent break-up outside of the inertial subrange if a more general model for the turbulent kinetic energy spectrum is available to estimate $u_{D_p}$ as a more involved function of $D_p$.

A complete characterization of locality requires knowledge of the break-up probability $q_b(D_c|D_p)$ as well. Compared to the scalings for $f$ and $g_b$ above, there is less consensus among analytical, experimental, and numerical studies on the appropriate scaling of $q_b$ with $D_c$ and $D_p$ in the context of turbulent break-up. Various model forms have been developed from statistical ansatzes, phenomenological arguments, and empirical data, as reviewed in detail by~\citet{Lasheras1},~\citet{Liao1},~\citet{MartinezBazan3}, and~\citet{Solsvik2}. Two canonical distributions in bubble-volume space are used as surrogate models to cover the range of these model forms: the uniform distribution and the beta distribution. The validity of these surrogate models will be examined using the simulations in Part 2.

\subsection{Uniform distribution in bubble-volume space}\label{sec:qbuniform}

Consider, first, the uniform distribution in bubble-volume space ($D^3$-space)
\begin{equation}
q_b\left(D_c^3|D_p^3\right) =
\begin{cases}
\f{2}{D_p^3}, &\qquad 0 \leq D_c^3 \leq D_p^3,\\
0, &\qquad D_p^3 < D_c^3,
\end{cases}
\end{equation}
where the factor of 2 arises from the assumption of binary break-up. From the properties of $q_b$ discussed in appendix~\ref{app:kernelmore}, this is equivalent to the following distribution in bubble-size space ($D$-space)
\begin{equation}
q_b(D_c|D_p) =
\begin{cases}
\f{6D_c^2}{D_p^3}, &\qquad 0 \leq D_c \leq D_p,\\
0, &\qquad D_p < D_c,
\end{cases}
\end{equation}
where the additional factor of 3 arises from the change in variables from $D^3$ to $D$. With the available scalings for $q_b$ and $g_b f$, the relations \eqref{eqn:WbIRI} and \eqref{eqn:WbUVI} yield
\begin{gather*}
I_p(D_p|D) \sim \int_0^D \diff D_c \: D_c^5 D_p^{-7} \sim D^6 D_p^{-7}, \qquad I_c(D_c|D) \sim \int_D^\infty \diff D_p \: D_c^5 D_p^{-7} \sim D_c^5 D^{-6}.
\end{gather*}
Observe that $I_p$ and $I_c$ rapidly decrease as $D_p\rightarrow \infty$ and $D_c\rightarrow 0$, respectively, indicating that $W_b$ may be reasonably approximated as size local. More specifically, the limits
\begin{gather}
I_p(D_p|D) \sim D_p^{\gamma_p} \sim D_p^{-7},\label{eqn:uniformIpinf}\\
I_c(D_c|D) \sim D_c^{\gamma_c} \sim D_c^{5}\label{eqn:uniformIczero}
\end{gather}
hold as $D_p \rightarrow \infty$ and $D_c \rightarrow 0$, respectively. The exponents $\gamma_p$ and $\gamma_c$ were referenced earlier in figures \ref{fig:IRlocality}(b) and \ref{fig:UVlocality}(b), respectively. Note that these relations hold even at $D_p \sim D$ and $D_c \sim D$, respectively, because $q_b$ is separable in $D_p$ and $D_c$. Thus, a stronger statement on locality may be made in the case of the uniform distribution: since
\begin{equation}
W_b(D) \sim \left(\int_0^D \diff D_c \: D_c^5 \right) \times \left( \int_D^\infty \diff D_p \: D_p^{-7}\right)
\end{equation}
may be expressed as the separable product of two integrals, one may further conclude that $W_b(D)$ may be directly approximated by a movement of bubble mass in bubble-size space from some bubble size just larger than $D$ to some bubble size just smaller than $D$. Finally, as a self-consistency check, one may obtain the scaling of $W_b(D)$ with $D$
\begin{equation}
W_b(D) \sim \int_0^D \diff D_c \: D_c^5 D^{-6} \sim \int_D^\infty \diff D_p \: D^6 D_p^{-7} \sim D^6 D^{-6} \sim \text{constant}.
\end{equation}
If the underlying energy flux in the surrounding turbulence is scale invariant within an inertial scale subrange, and the break-up probability is chosen to be size invariant in a corresponding intermediate range of bubble sizes, then the resulting bubble break-up flux is size invariant, confirming the presence of an intermediate size subrange where the break-up process is self-similar in nature. Self-similarity is compatible with the assumption of statistical quasi-stationarity and quasi-homogeneity as evidenced by \eqref{eqn:pbe_fbar2} and \eqref{eqn:WbTb}, assuming only $T_b$ is active on the right-hand side of \eqref{eqn:pbe_fbar2} in this intermediate size subrange. The potential non-stationarity of a non-self-similar break-up process is further addressed in appendix~\ref{app:caveats-self-similar}.

\subsection{Beta distribution in bubble-volume space}\label{sec:qbbeta}

\begin{figure}
  \centerline{
\includegraphics[width=0.7\linewidth]{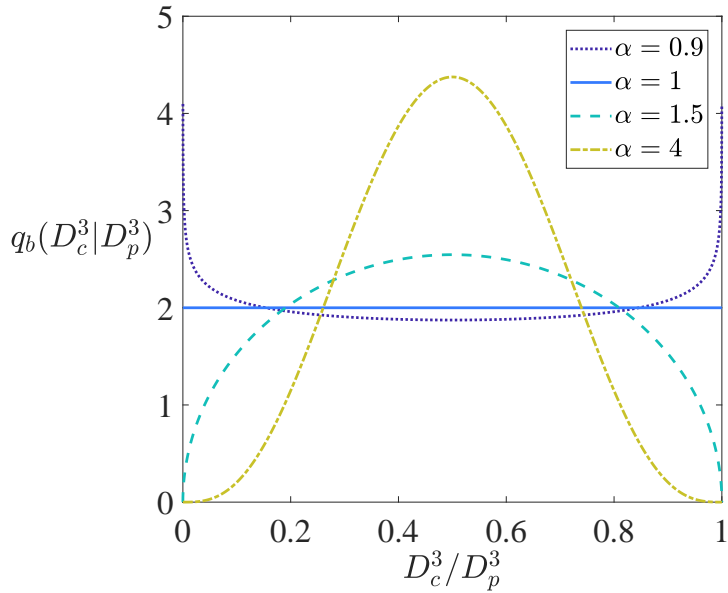}
}
  \caption{The symmetric beta distribution in $D^3$-space \eqref{eqn:betadis} for various shape parameters $\alpha$.} 
\label{fig:beta}
\end{figure}

\begin{figure}
  \centerline{
(a)
\includegraphics[width=0.7\linewidth,valign=t]{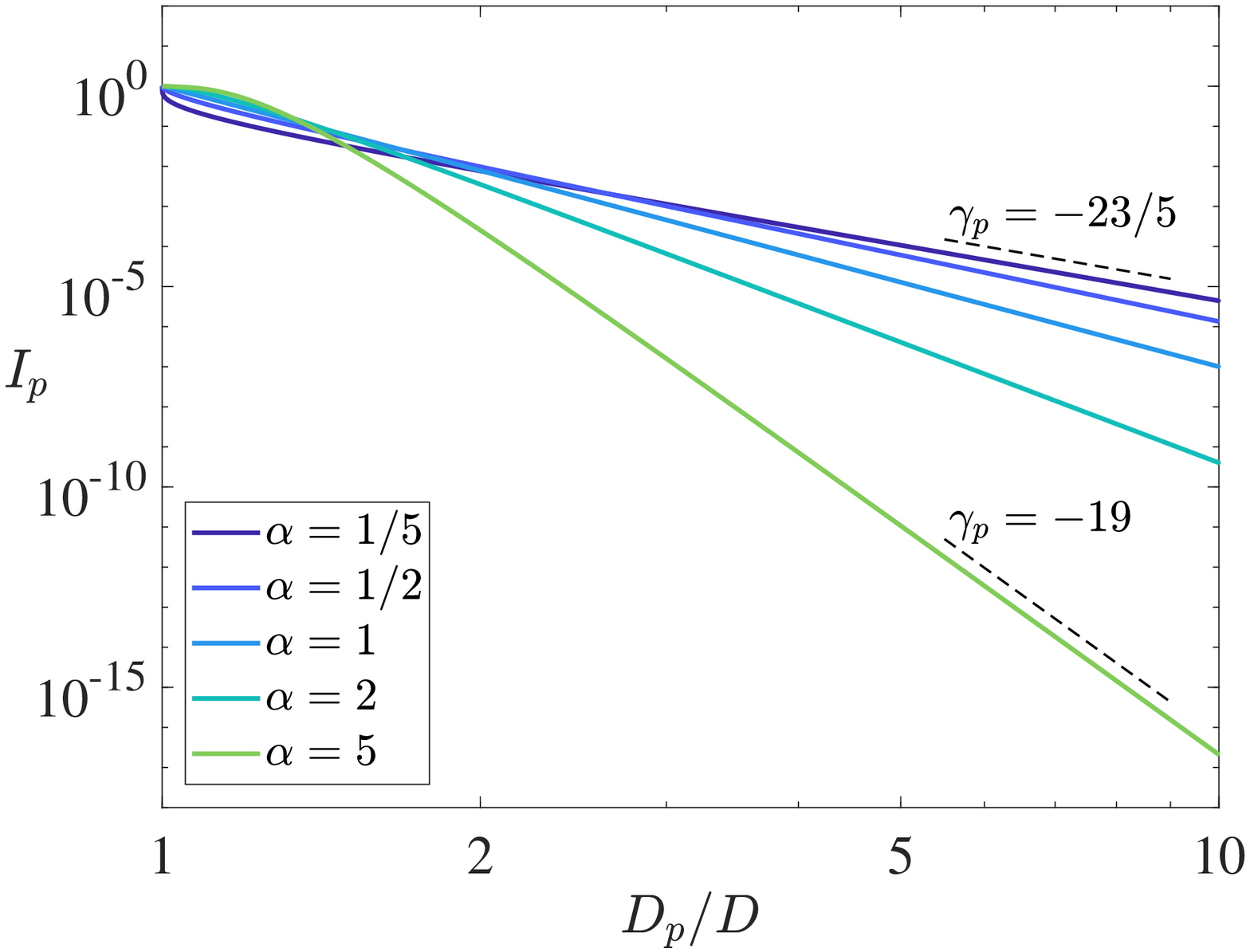}
}
  \centerline{
(b)
\includegraphics[width=0.7\linewidth,valign=t]{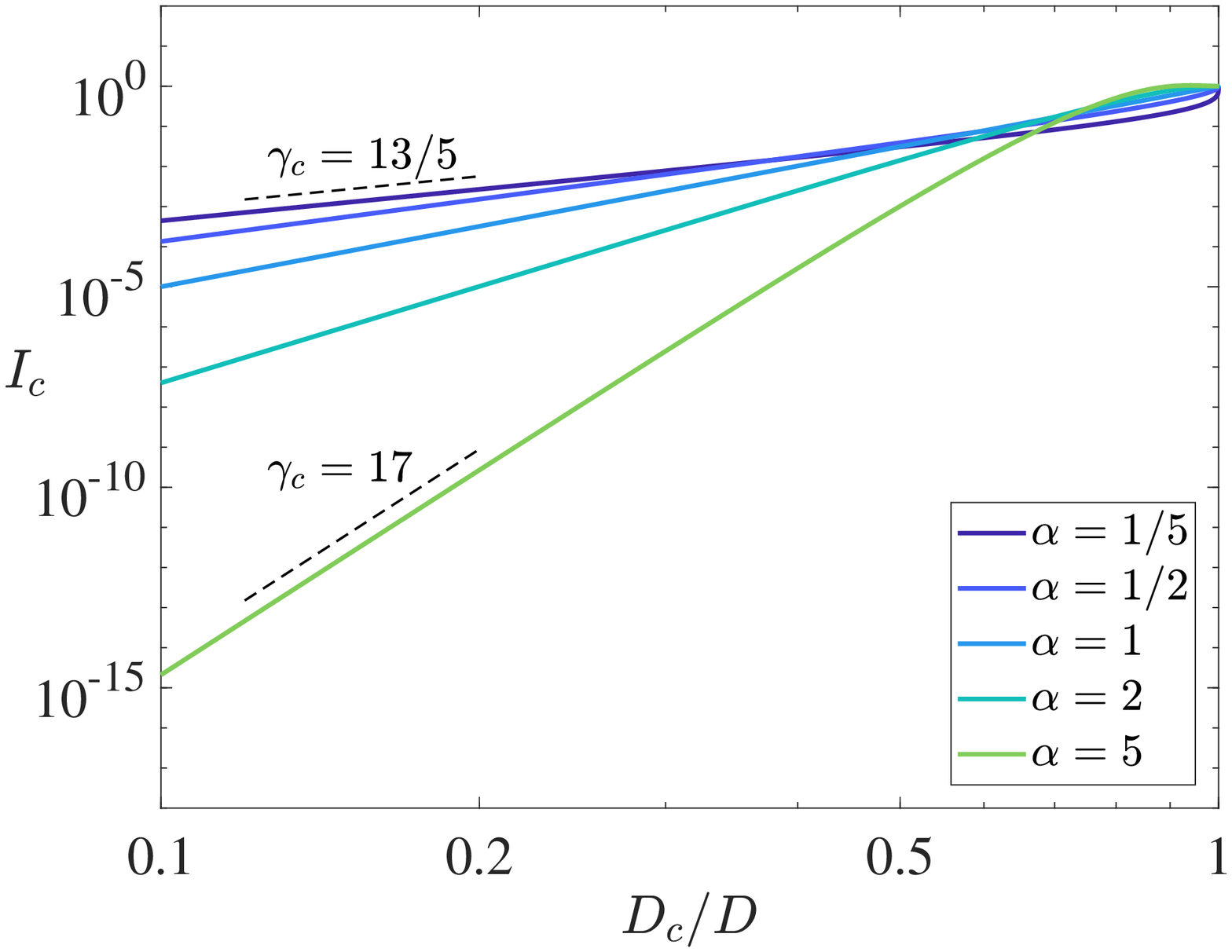}
}
  \caption{Integrands of $W_b$ demonstrating the extent of (a) infrared locality using \eqref{eqn:betaIR} and (b) ultraviolet locality using \eqref{eqn:betaUV} after substituting the symmetric beta distribution with various shape parameters $\alpha$ for the probability distribution of child bubble volumes $q_b$, as well as scalings for the differential bubble break-up rate $g_b f$ corresponding to a turbulent break-up mechanism. Since the proportionality constants are dropped in \eqref{eqn:betaIR} and \eqref{eqn:betaUV}, the integrands here are plotted in arbitrary units, with the values at $D_p = D$ [for (a)] and $D_c = D$ [for (b)] fixed at unity. The power-law fits at large $D_p/D$ and small $D_c/D$ correspond to the scaling limits in \eqref{eqn:betaIpinf} and \eqref{eqn:betaIczero} for $\alpha=1/5$ and $\alpha=5$.} 
\label{fig:betalocal}
\end{figure}

Recall from \S~\ref{sec:kernel} that the break-up probability is symmetric in bubble-volume space. The beta distribution that satisfies this constraint can take only a single shape parameter $\alpha$, and may be expressed in bubble-volume space, or $D^3$-space, as
\begin{equation}
q_b\left(D_c^3|D_p^3\right) =
\begin{cases}
2D_c^{3(\alpha-1)}\left(D_p^3-D_c^3\right)^{\alpha-1}D_p^{-6(\alpha-1)-3}/B(\alpha,\alpha), &\qquad 0 \leq D_c^3 \leq D_p^3,\\
0, &\qquad D_p^3 < D_c^3,
\end{cases}
\label{eqn:betadis}
\end{equation}
where $B(\alpha,\alpha)$ is the beta function~\citep[\S~6.2]{Abramowitz1}, which is a normalization constant for the beta distribution with shape parameter $\alpha$, and the factor of 2 arises from the assumption of binary break-up. This distribution is plotted in figure~\ref{fig:beta} for several values of $\alpha$. Note that the uniform distribution is recovered when $\alpha=1$. The beta distribution is defined only for $\alpha > 0$. When $0 < \alpha < 1$, the distribution is U-shaped and goes to infinity at the endpoints of the domain, thus favouring the formation of bubbles of unequal sizes. The formation of these bubbles is permitted in the infinite-Weber-number limit where the Hinze scale is zero, as discussed in the introduction. For practical flows with finite integral-scale Weber numbers, the favourable formation of bubbles of sizes smaller than the Hinze scale may not be as plausible, and a more precise surrogate model for $q_b$ may have to involve a truncated U-shaped beta distribution, or an M-shaped distribution. Nevertheless, the U-shaped beta distribution should remain an adequate surrogate model for parent bubbles of sizes $L_n$ in the intermediate size subrange $L_\mathrm{H} \ll L_n \ll L$ and sufficiently larger than the Hinze scale. When $\alpha > 1$, the distribution is inverted-U-shaped and goes to zero at the endpoints, thus favouring the formation of bubbles of equal sizes. The beta distribution is thus a reasonable surrogate model for most observed and modelled break-up distributions, except for the class of M-shaped distributions. One such distribution was introduced by~\citet{Wang2}; see the reviews cited in the preamble of this section for more examples. The analogue of \eqref{eqn:betadis} in bubble-size space, or $D$-space, is
\begin{equation}
q_b(D_c|D_p) =
\begin{cases}
6D_c^{3\alpha-1}\left(D_p^3-D_c^3\right)^{\alpha-1}D_p^{3-6\alpha}/B(\alpha,\alpha), &\qquad 0 \leq D_c^3 \leq D_p^3,\\
0, &\qquad D_p^3 < D_c^3.
\end{cases}
\end{equation}
With the available scalings for $q_b$ and $g_b f$, the relations \eqref{eqn:WbIRI} and \eqref{eqn:WbUVI} yield
\begin{gather}
I_p(D_p|D) \sim \int_0^D \diff D_c \: \f{D_c^{3\alpha+2}}{\left(D_p^3-D_c^3\right)^{1-\alpha}}D_p^{-1-6\alpha} \sim D_p^{-1} \int_0^{D^3/D_p^3} \diff x \: x^\alpha (1-x)^{\alpha-1},\label{eqn:betaIR}\\
I_c(D_c|D) \sim \int_D^\infty \diff D_p \: \f{D_p^{-1-6\alpha}}{\left(D_p^3-D_c^3\right)^{1-\alpha}} D_c^{3\alpha+2} \sim D_c^{-1} \int_0^{D_c^3/D^3} \diff x \: x^\alpha (1-x)^{\alpha-1}.\label{eqn:betaUV}
\end{gather}
The final integrals in \eqref{eqn:betaIR} and \eqref{eqn:betaUV} are the incomplete beta functions $B_{D^3/D_p^3}(\alpha+1,\alpha)$ and $B_{D_c^3/D^3}(\alpha+1,\alpha)$, respectively~\citep[\S~6.6.1 and 26.5.3]{Abramowitz1}. The results discussed in \S~\ref{sec:qbuniform} are exactly recovered when $\alpha=1$. The expressions in \eqref{eqn:betaIR} and \eqref{eqn:betaUV} are plotted in arbitrary units as functions of $D_p/D$ and $D_c/D$, respectively, in figure~\ref{fig:betalocal}. As $\alpha$ increases, the rates of decay of $I_p$ and $I_c$ as $D_p\rightarrow\infty$ and $D_c\rightarrow 0$, respectively, increase, indicating that as break-up events involving children bubbles of similar sizes are increasingly favoured, the locality of the break-up process correspondingly increases. At small $\alpha$, where the most likely break-up events involve children bubbles of very different sizes, the cascade is diffuse, or leaky, and the bubble break-up flux is less local. Also, for sufficiently large $D_p$ and sufficiently small $D_c$, \eqref{eqn:betaIR} and \eqref{eqn:betaUV} may respectively be approximated as
\begin{gather}
I_p(D_p|D) \approx \f{D_p^{-1-6\alpha}}{D_p^{3(1-\alpha)}} \sim D_p^{-4-3\alpha} \sim D_p^{\gamma_p},\label{eqn:betaIpinf}\\
I_c(D_c|D) \approx D_c^{3\alpha+2} \sim D_c^{\gamma_c}.\label{eqn:betaIczero}
\end{gather}
Recall that the exponents $\gamma_p$ and $\gamma_c$ were referenced earlier in figures \ref{fig:IRlocality}(b) and \ref{fig:UVlocality}(b), respectively. Note, also, that in these limits, $I_p$ decays at least as quickly as $D_p^{-4}$, and $I_c$ grows at least as quickly as $D_c^2$, so the break-up flux is always at least quasi-local regardless of $\alpha$, for values of $\alpha$ where the beta distribution is defined. Once again, the results of \S~\ref{sec:qbuniform} are recovered\textemdash in an exact fashion\textemdash for $\alpha=1$. Note, in addition, that a bubble break-up process best described by an M-shaped distribution may be modelled by a superposition of two bubble-mass fluxes due to two $q_b$'s with different $\alpha$'s. Since each bubble-mass flux is always at least quasi-local regardless of $\alpha$, this implies that M-shaped distributions also result in a net quasi-local flux. Finally, one may also examine the dependence of $W_b(D)$ on $D$ as a self-consistency check
\begin{align}
W_b(D) &\sim \int_0^D \diff D_c \: \left[ D_c^{-1} \int_0^{D_c^3/D^3} \diff x \: x^\alpha (1-x)^{\alpha-1} \right] \notag\\
&\sim \int_D^\infty \diff D_p \: \left[ D_p^{-1} \int_0^{D^3/D_p^3} \diff x \: x^\alpha (1-x)^{\alpha-1} \right] \notag\\
&\sim \int_0^1 \diff y \: \left[ y^{-1} \int_0^{y^3} \diff x \: x^\alpha (1-x)^{\alpha-1} \right] \notag\\
&\sim \text{constant},
\end{align}
which reveals, as expected, an intermediate bubble-size subrange for the bubble break-up flux where the break-up process is self-similar in nature, if $\alpha$ is constant over the size subrange. Once again, self-similar behaviour of $W_b$ is compatible with the statistical quasi-stationarity and quasi-homogeneity of the system $\left(T_b = 0\right)$. The potential non-stationarity of a non-self-similar break-up process is further addressed in appendix~\ref{app:caveats-self-similar}.

\subsection{Revisiting some of the assumptions in the bubble break-up formalism}

Note that the findings of this work assume that all break-up events are binary in nature. The binary break-up assumption precludes the formation of satellite bubbles, which might be assumed to decrease the locality of the resulting bubble-mass transfer and also disrupt self-similarity. It turns out, however, that locality and self-similarity remain plausible in such a scenario. Non-binary break-up events are addressed in appendix~\ref{app:caveats-binary}. As mentioned earlier, appendix~\ref{app:caveats-self-similar} discusses non-self-similar break-up mechanisms, which may be relevant in systems without sufficient scale separation. It turns out that locality remains relatively robust even in the absence of self-similarity.

The extent of locality in the break-up flux $W_b(D)$, in particular the respective scalings of $I_p(D_p|D)$ and $I_c(D_c|D)$ with $D_p$ and $D_c$, will be examined in Part 2 via a direct evaluation of the flux from all relevant break-up events in a breaking-wave simulation.


\section{Model descriptions for bubble-mass transfer and their implications on subgrid-scale modelling}\label{sec:model}

\subsection{Relations between bubble-mass and spectral energy flux models}\label{sec:spectralenergy}

The break-up flux $W_b(D) = \int_0^D \diff D_c \: T_b(D_c)$ describes the average movement of bubble mass $\sim f(D)D^3$ in bubble-size space ($D$-space) as governed by the population balance equation given in \eqref{eqn:pbe_fbar} and \eqref{eqn:pbe_fbar2}, where the rate of change of $f(D)D^3$ due to break-up is $T_b(D)$, recalling from the introduction that mass and volume are assumed to be equivalent in the incompressible limit. This is analogous to how the transfer flux $W(k) = \int_0^k \diff k' \: T(k')$ describes the movement of turbulent kinetic energy $E(k)$ in wavenumber space ($k$-space) based on the spectral turbulent kinetic energy equation~\citep{Batchelor1}
\begin{equation}
\f{\partial E(k,t)}{\partial t} = T(k,t) - 2\nu_l k^2 E(k,t),
\end{equation}
where the rate of change of $E(k)$ due to interscale transfer is $T(k)$, and the time dependence drops off in the quasi-stationary limit. In particular, the double-integral form of the break-up flux \eqref{eqn:Wb} is reminiscent of the spectral energy transfer model of~\citet{Heisenberg1,Heisenberg2}, where $W(k)$ is modelled as a separable product of integrals
\begin{equation}
W(k) \sim \left( \int_0^k \diff k' \: E(k') {k'}^2 \right) \times \left( \int_k^\infty \diff k'' \: \sqrt{\f{E(k'')}{{k''}^3}} \right).
\label{eqn:heisenberg}
\end{equation}
By substituting the inertial subrange scaling $E(k) \sim k^{-5/3}$ into \eqref{eqn:heisenberg}, one obtains ${k'}^{-5/3} {k'}^2 \sim {k'}^{1/3}$ and $\sqrt{{k''}^{-5/3}{k''}^{-3}} \sim {k''}^{-7/3}$ for the scalings of the two integrands, suggesting some degree of infrared and ultraviolet locality, respectively. Remarkably, it turns out that these model limits agree with the scalings obtained in the analyses by \citet{Eyink1}, \citet{Eyink2}, and \citet{Aluie1} for the turbulent energy cascade for a monofractal velocity field, as well as an earlier analysis by \citet{Kraichnan1} based on closure approximations that also introduces a measure of locality, and earlier investigations by~\citet{Zhou1,Zhou2} based on numerical simulations. Note that these rates of decay are slower than those obtained in \S~\ref{sec:qbuniform} and \S~\ref{sec:qbbeta} for the bubble-mass flux integrands, suggesting that the turbulent bubble-mass cascade may be more strongly local than the turbulent energy cascade. One may further evaluate these integrals
\begin{equation}
W(k) \sim \int_0^k \diff k' \: k'^{1/3} \int_k^\infty \diff k'' \: {k''}^{-7/3} \sim k^{4/3}k^{-4/3} \sim \text{constant},
\end{equation}
in order to see that $W(k)$ has no dependence on $k$, as one would expect for a self-similar energy transfer process. Note that this self-similarity, in turn, implies that the underlying system dynamics are statistically steady or quasi-stationary, since $T(k)$ must then be negligible in the range of scales of interest. An analogous observation was made in the case of the bubble break-up flux in \S~\ref{sec:qbuniform} and \S~\ref{sec:qbbeta}.

If the true $W(k)$ is quasi-local in $k$-space, then it may be well approximated by a wavenumber-local expression. This brings to mind the quasi-local models of~\citet{Kovasznay1} and~\citet{Pao1,Pao2}. \citet{Kovasznay1} argued that if $W(k)$ is dependent only on $E(k)$ and $k$, then the only dimensionally consistent expression is
\begin{equation}
W(k) \sim \left[E(k)\right]^{3/2} k^{5/2}.
\end{equation}
Subsequently, \citet{Pao1,Pao2} allowed $W(k)$ to depend on $\varepsilon$ as well. If it is further assumed that $W(k)$ is linear in $E(k)$, then it follows from dimensional arguments that
\begin{equation}
W(k) \sim \varepsilon^{1/3} k^{5/3} E(k).
\end{equation}
In a similar fashion, $W_b(D)$ may be justifiably modelled by an expression local in $D$-space if there is sufficient quasi-locality in the break-up flux. From a comparison of \eqref{eqn:pbe_locality} and \eqref{eqn:WbTb}, it is clear that the local transport term in the phase-space-based population balance equation provides an appropriate model form for a local $W_b$. Then, one desires an appropriate model for the velocity of $f(D)D^3$ in bubble-size space, $v_D(D)$, such that 
\begin{equation}
W_b(D) \sim v_D(D) f(D) D^3.
\label{eqn:localflux}
\end{equation}
If there exists an intermediate bubble-size subrange where $W_b(D)$ is independent of $D$, and $f(D) \sim D^{-10/3}$, then an appropriate model for $v_D(D)$ should satisfy 
\begin{equation}
v_D(D) \sim D^{1/3}.
\end{equation}
Note that this is similar to the scaling for the turbulent velocity fluctuations with eddy size $u_D(D) \sim D^{1/3}$. The scaling for $v_D$ was previously postulated by~\citet{Garrett1} on the dimensional grounds that $v_D \sim u_D$, but one should be cognizant of the difference between bubble-size space and eddy-size space. In addition, the term $\partial \left(v_D f D^3\right)/\partial D$ in the original supporting reference~\citep{Garrettson1} was used to model a dissolution process, meaning that the model form referenced by~\citet{Garrett1} is applicable only to the change in bubble mass in individual events. Here, the model form for size-local bubble-mass transport is not applicable to individual events, as will be clarified by the discussion in appendix~\ref{app:indivevents}. The locality of the corresponding bubble-mass flux must necessarily be interpreted in an averaged sense, as all turbulent cascades should be. In turn, the model velocity $v_D$ strictly describes the \emph{averaged} break-up dynamics in small, localized regions of turbulent bubbly flows with sufficient scale separation.

\begin{figure}
  \centerline{
\includegraphics[width=0.7\linewidth]{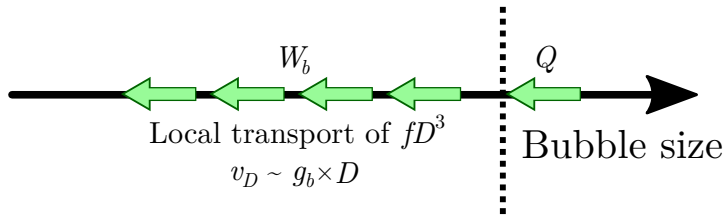}
}
  \caption{Schematic of local transport of bubble mass $\sim fD^3$ in $D$-space. The large-scale entrainment rate $Q$ is directly related to the bubble-mass flux $W_b \sim v_D f D^3 \sim g_b f D^4$.} 
\label{fig:pbe_local}
\end{figure}

To close this discussion, recall the scaling for the break-up frequency $g_b(D) \sim D^{-2/3}$, which may be interpreted as the inverse of the characteristic break-up time of bubbles of size $D$. If one assumes that the flux $W_b(D)$ is effectively described by a size-space velocity $v_D(D)$ such that a characteristic size interval $D$ is travelled in this characteristic time, then one may write $v_D(D) \sim g_b(D)D$, and thus $W_b(D) \sim g_b(D)f(D)D^4$. The scaling $g_b(D) f(D) \sim D^{-4}$ is thus seen to follow directly from the assumption of a quasi-local and self-similar bubble break-up flux. The scaling of~\citet{Garrett1} for $f \sim Q \varepsilon^{-1/3} D^{-10/3}$, obtained via dimensional analysis in an intermediate bubble-size subrange using a steady large-scale entrainment rate $Q$, is also a direct consequence of quasi-locality and self-similarity in the bubble-mass flux, with the additional consideration that $Q \sim W_b$. This exhibits a clear parallel to the turbulent energy cascade, where it is also typically assumed that the energy production and cascade rates are of the same order of magnitude. Some of these ideas are summarized in the schematic on bubble-mass transport in figure~\ref{fig:pbe_local}. Further remarks on $Q$ are provided in appendix~\ref{app:entrainment}.

\subsection{Implications for subgrid-scale modelling}\label{sec:SGS}

\begin{figure}
  \centerline{
\includegraphics[width=0.7\linewidth]{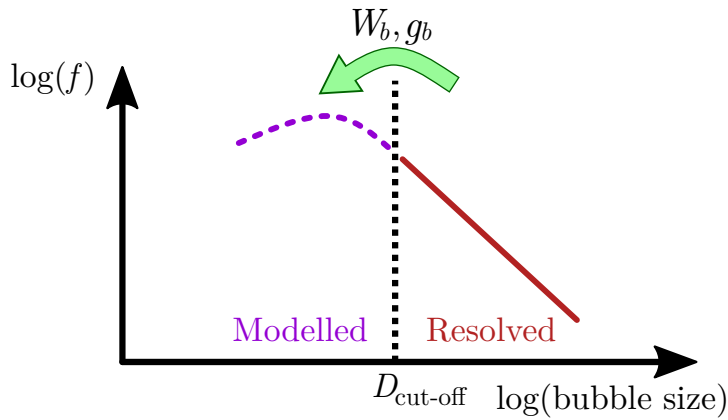}
}
  \caption{Schematic illustrating subgrid-scale modelling in turbulent bubbly flows.} 
\label{fig:SGSmodel}
\end{figure}

Aside from providing a theoretical basis for the scalings for $f$ and $v_D$ proposed by~\citet{Garrett1} through connections to the characteristic break-up frequency $g_b$~\citep{Kolmogorov3,Hinze1,MartinezBazan1}, this work has also posited that the bubble break-up cascade provides a universal description of the bubble break-up dynamics at small and intermediate bubble sizes in small, localized regions of turbulent bubbly flows with sufficient scale separation. For example, the break-up flux in these small, localized regions should be constant in an intermediate subrange of bubble sizes $L_\mathrm{H} \ll L_n \ll L$, provided the surrounding turbulence is sufficiently energetic. Universality simplifies the task of subgrid-scale modelling in turbulent two-phase flows with a large separation of scales, and lends legitimacy to a universal subgrid-scale model in the spirit of LES of turbulent single-phase flows. In traditional LES, large-scale turbulent motions and flow structures are resolved, while small-scale motions and structures are modelled. The rationale for this approach is two-fold, as discussed succinctly by~\citet{Rogallo1}. Large-scale motions are influenced by the flow geometry and cannot be assumed to have a universal character. They are thus explicitly resolved, along with the bulk of the energy in the flow. Small-scale motions may be assumed to have a universal character and are instead represented by models that dissipate energy in a universal fashion. A similar idea may be applied to turbulent two-phase flows where a separation of scales enables a universal description of the small scales. Large structures of the dispersed phase are explicitly resolved via an interface-tracking or interface-capturing method, while small structures of the dispersed phase are treated as subgrid entities using a Lagrangian point-particle description. If the formation and dynamics of these subgrid bubbles occur in a universal fashion, then simplified models may be used to generate these bubbles through the modelled break-up of larger bubbles. For example, the results of this work suggest that in simulations where the mesh resolution is larger than the expected $L_\mathrm{H}$, the generation of super-Hinze-scale subgrid bubbles may be modelled via a bubble break-up cascade, as illustrated in figure~\ref{fig:SGSmodel}. As noted in the introduction, most sub-Hinze-scale bubbles are expected to be formed by distinct fragmentation mechanisms, such as Mesler entrainment, as well as regular and irregular drop entrainment~\citep{Deane1,Kiger1,Chan3,Chan6}. As such, the generation of sub-Hinze-scale subgrid bubbles will have to be addressed separately in a manner that bypasses the cascade considered in this work~\citep[see, e.g.,][]{Chan5,Chan3,Chan6}. It is envisioned that distinct subgrid-scale models for sub-Hinze-scale and super-Hinze-scale subgrid bubbles be combined in an additive fashion in order to account for this myriad of fragmentation mechanisms and cover more bases for modelling the formation and dynamics of subgrid bubbles. A detailed formulation of a suitable subgrid-scale model in the context of super-Hinze-scale subgrid bubbles is under development. This model would use both the kernel-based break-up model form~\eqref{eqn:Wb}, as well as the phase-space-based break-up model form~\eqref{eqn:localflux}. 

\section{Conclusions}\label{sec:conclusions}

This paper explores the properties of the bubble break-up cascade that was postulated by~\citet{Garrett1} to generate a spectrum of bubble sizes beneath breaking waves, and more generally in high-Reynolds-number and high-Weber-number turbulent flows. The description of the turbulent bubble-mass cascade is strongly analogous to the turbulent energy cascade in single-phase turbulence~\citep{Richardson1,Kolmogorov1,Onsager1}. An intrinsic feature of these cascades is the approximate scale locality of interscale fluxes. In the case of the bubble-mass cascade, this specifically refers to the bubble-mass flux from large to small bubble sizes, which is governed by bubble break-up event statistics. Novel manipulation of a mass-conserving population balance equation for the bubble size distribution, $f$, is shown to yield quantitative insights into the locality of this flux. The key ingredient for locality is the adoption of turbulent-flow scalings for $f$ and the break-up frequency, $g_b$, with theoretical, numerical, and experimental support. With these scalings, the flux is shown to be infrared local, where flux contributions from parent bubbles of sizes $D_p > D$ decay faster than $(D_p/D)^{-4}$, and ultraviolet local, where flux contributions to children bubbles of sizes $D_c < D$ decay faster than $(D/D_c)^{-2}$. In other words, the bubble-mass flux is approximately size local with a power-law decay for longer-range interactions. These flux scalings suggest that the turbulent bubble-mass cascade is more strongly local than the turbulent energy cascade. The presence of locality is not too sensitive to the probability distribution of child bubble volumes, $q_b$, but the shape of the distribution influences the strength of locality. In the case of the uniform distribution, for example, flux contributions from parent bubbles may decay as quickly as $(D_p/D)^{-7}$, and flux contributions to children bubbles may decay as quickly as $(D/D_c)^{-5}$. Under the assumptions of quasi-stationarity and quasi-homogeneity, it may be further deduced that the bubble break-up flux is self-similar in an intermediate bubble-size subrange, much like the energy flux in the inertial subrange in the energy cascade. The theoretical tools introduced here in Part 1 enable detailed inspection of numerical simulations of breaking waves in a forthcoming companion paper, Part 2, through a detailed analysis of bubble break-up statistics. Taken together, these findings confirm key physical aspects of the turbulent bubble break-up cascade phenomenology and provide a theoretical basis for the dimensional analysis of~\citet{Garrett1} using traditional turbulent-flow scalings for bubble break-up~\citep{Kolmogorov3,Hinze1,MartinezBazan1}. Locality in the bubble-mass transfer process implies that small-bubble break-up may be universal in small, localized regions in a variety of turbulent bubbly flows with sufficient scale separation. In particular, the results of this work have not been specifically derived for oceanic breaking waves, and might be broadly applicable to other turbulent two-phase flows under appropriate conditions, such as bubble break-up in stirred tanks and reactors. This universality lends legitimacy to the construction of universal subgrid-scale models for the break-up of subgrid bubbles in LES of these flows.

On average, the bubble break-up cascade transfers bubble mass from large to small bubble sizes. The sustained presence of this break-up cascade implies the eventual dominance of bubble dynamics by these small bubbles. Small bubbles are known to linger in terrestrial air--water flows due to their low rise velocity~\citep{Garrettson1,Thorpe2,Thorpe1,Trevorrow1}. Knowledge of the behaviour of these bubbles is thus of practical importance for characterizing these flows. Effective predictive modelling of the statistics of these bubbles leads to accurate prediction of physical phenomena related to the acoustical and optical responses of these bubbles, such as the persistent wake signatures of seafaring vessels. The results of Part 2 will demonstrate the relevance of this cascade mechanism in realistic air--water flow configurations, while the modelling approach to be introduced in forthcoming work is a step towards accurate physics-based prediction of small-bubble statistics in these practical configurations.

\section*{Acknowledgments} 

This investigation was funded by the Office of Naval Research, Grant \#N00014-15-1-2726, and is also supported by the Advanced Simulation and Computing programme of the U.S. Department of Energy’s National Nuclear Security Administration via the PSAAP-II Center at Stanford University, Grant \#DE-NA0002373. W.~H.~R. Chan is also funded by a National Science Scholarship from the Agency of Science, Technology and Research in Singapore. The authors acknowledge computational resources from the U.S. Department of Energy's INCITE Program. The authors would like to thank J. Urzay, A. Mani, D. Livescu, A. Lozano-Dur\'{a}n, and M.~S. Dodd for useful discussions, as well as S.~S. Jain and H. Hwang for their comments on an early version of this manuscript.

\section*{Declarations}

The authors report no conflict of interest.

\appendix

\section{Similarity hypotheses for the turbulent energy and bubble-mass cascades}\label{app:similarity}

\subsection{Kolmogorov's similarity hypotheses for high-Re single-phase turbulent flows}\label{app:energyhypo}

Kolmogorov's hypotheses for the local structure of turbulence in high-$\RR_L$ flows~\citep{Kolmogorov1}, which were phenomenologically reviewed in detail in \S~\ref{sec:energycascade}, were paraphrased by~\citet[\S~6.1.2]{Pope1} and are further paraphrased here for reference:

\begin{hypothesis}\label{hyp:localisotropy}
\normalfont
(Local isotropy) In flows with sufficiently high Reynolds number, the small-scale turbulent motions $(L_n \ll L)$ are isotropic.
\end{hypothesis}

This hypothesis echoes the statements earlier that scalings relevant to statistically stationary and homogeneous turbulent flows also apply in small, localized regions in a variety of turbulent flows with sufficient scale separation.

\begin{hypothesis}
\normalfont
(First similarity hypothesis) For locally isotropic turbulence, the statistics of the small-scale turbulent motions $(L_n \ll L)$ have a universal form that is uniquely determined by $\varepsilon$ and $\nu_l$.
\end{hypothesis} 

\begin{hypothesis}\label{hyp:inertialsubrange}
\normalfont
(Second similarity hypothesis) At scales in the range $L_\mathrm{K} \ll L_n \ll L$ in locally isotropic turbulence, the statistics of the turbulent motions have a universal form that is uniquely determined by $\varepsilon$ and independent of $\nu_l$.
\end{hypothesis} 

Note that in the original hypotheses, the ``statistics of the turbulent motions" refer specifically to the statistics of the second-order velocity structure functions.

\subsection{Proposed similarity hypotheses for high-Re and high-We turbulent bubbly flows}\label{app:bubblehypo}

A corresponding set of similarity hypotheses pertaining to the turbulent bubble-mass cascade examined in \S~\ref{sec:bubblecascade} is proposed here:

\begin{hypothesis}\label{hyp:sphericity}
\normalfont
(Single-size approximation) In bubbly flows with sufficiently high Weber number, the statistics of sufficiently small bubbles of volumes $L_n^3 \ll L^3$ may be analyzed by parameterizing each bubble by a single length scale $L_n$.
\end{hypothesis}

If the phase space of the bubble size distribution contains no other important dimensions, then the single-size approximation enables the treatment of the distribution as a one-dimensional probability distribution in bubble-size space.

\begin{hypothesis}\label{hyp:bubbleuniversality}
\normalfont
(First similarity hypothesis for gas transfer in bubble-size space due to turbulent break-up) The statistics of sufficiently small bubbles of sizes $L_n \ll L$ have a universal form that is uniquely determined by $\varepsilon$ and $\sigma/\rho_l$.
\end{hypothesis} 

\begin{hypothesis}\label{hyp:intermediatesubrange}
\normalfont
(Second similarity hypothesis for gas transfer in bubble-size space due to turbulent break-up) The statistics of bubbles of sizes $L_\mathrm{H} \ll L_n \ll L$ have a universal form that is uniquely determined by $\varepsilon$ and independent of $\sigma/\rho_l$.
\end{hypothesis} 

Hypothesis~\ref{hyp:intermediatesubrange} implies the presence of an intermediate bubble-size subrange for bubble-mass transfer in turbulent bubbly flows with sufficiently high $\We_L$, in an analogous fashion to the inertial subrange implied by hypothesis~\ref{hyp:inertialsubrange}. 

The proposed hypotheses for the turbulent bubble-mass cascade are chiefly applicable to low-order bubble statistics like the bubble size distribution $f$, by analogy with the low-order flow statistics referenced by Kolmogorov's original similarity hypotheses.


\section{Contributions of individual break-up events to the break-up flux $W_b$}\label{app:indivevents}

\begin{figure}
  \centerline{
\includegraphics[width=0.7\linewidth]{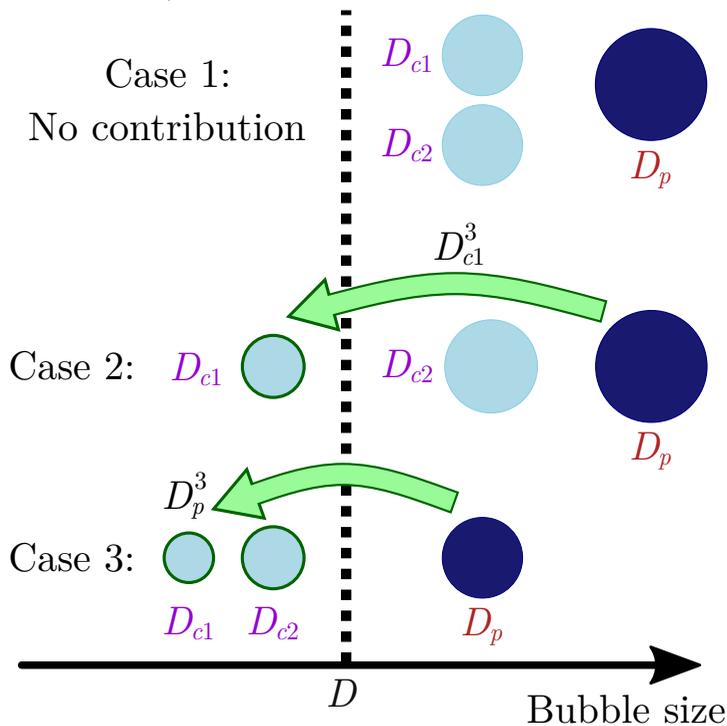}
}
  \caption{Schematics illustrating the three cases of break-up events discussed in appendix~\ref{app:indivevents}. Parent bubbles have a dark fill colour, while children bubbles have a light fill colour. Children bubbles that contribute to the bubble-mass flux $W_b(D)$ are marked with a dark border.} 
\label{fig:indivevents}
\end{figure}

The break-up flux $W_b$ introduced in \S~\ref{sec:kernel} is a statistical quantification of the bubble-mass transfer rate due to many independent break-up events, obtained through the ensemble-averaging operation discussed in \S~\ref{sec:sizedist}. The meaning and significance of locality may be elucidated by isolating the contributions of each break-up event to the flux $W_b(D)$. Assume that a parent bubble of size $D_p > D$ breaks up into two children bubbles of sizes $D_{c1}$ and $D_{c2}$. Non-binary break-up events are further discussed in appendix~\ref{app:caveats-binary}. By the conservation of mass, these bubble sizes must satisfy the constraint $D_p^3 = D_{c1}^3 + D_{c2}^3$. The contribution of each of these break-up events to the total flux $W_b(D)$ across the bubble size $D$ depends on the magnitudes of $D_{c1}$ and $D_{c2}$ relative to $D$. Any such break-up event has three possible outcomes. First, if $D_{c1}$ and $D_{c2}$ are both larger than $D$, then no bubble mass is transferred to any bubbles of sizes smaller than $D$. The resulting contribution to $W_b(D)$ in this case is zero. Second, if $D_{c1}$ is smaller than $D$ while $D_{c2}$ is larger than $D$, then the volume $D_{c1}^3$ is transferred from a bubble of size larger than $D$, i.e., $D_p$, to a bubble of size smaller than $D$, i.e., $D_{c1}$. Third, if both $D_{c1}$ and $D_{c2}$ are smaller than $D$, then the volume $D_p^3 = D_{c1}^3 + D_{c2}^3$ is transferred from a bubble of size larger than $D$ to bubbles of sizes smaller than $D$. These three cases are schematically illustrated in figure~\ref{fig:indivevents}. The relative frequency of these three cases is encapsulated in the break-up probability distribution $q_b(D_c|D_p)$ over all child bubble sizes $D_c \leq D_p$, as well as the ratio $D_p/D$. The average contribution of a single break-up event involving a parent bubble of size $D_p$ to the total flux $W_b(D)$ may be obtained by integrating the differential average volume transfer $q_b(D_c|D_p) D_c^3$ over all eligible child bubble sizes $D_c<D$. If these break-up events are independent of one another, then the total flux $W_b(D)$ may be constructed by multiplying this average gaseous volume transfer due to a single event involving a parent bubble of size $D_p$ by the corresponding differential event rate per unit domain volume $g_b(D_p)f(D_p)$, and then integrating over all eligible parent bubble sizes $D_p>D$. One may heuristically construct the expression~\eqref{eqn:Wb} for $W_b$ given these considerations.

Two observations may be made about the relationship between the bubble-mass transfer rate due to individual break-up events and the average flux $W_b$. First, an individual event that is itself non-local in bubble-size space may not contribute strongly to the non-locality of the corresponding $W_b$ if the frequency of this event is small. The intuition provided by a single event may thus not offer the complete story on the locality of $W_b$. Second, the bubble-mass transfer rate is a volume-weighted quantity. Consider the case where a parent bubble of size $D_p>D$ breaks into two children bubbles of sizes $D_{c1}<D$ and $D_{c2} \gg D_{c1}$. While this may appear to be a highly non-local event since $D_{c1}$ is far removed from $D_p$, the gaseous volume that is transferred from the bubble of size $D_p$ to the bubble of size $D_{c1}$ is $D_{c1}^3 \ll D_p^3$. The influence of this non-local transfer on the non-locality of $W_b(D)$ is limited by this volume weighting.

\section{More about the mathematical formalism}\label{app:formalismmore}

\subsection{The population balance equation}\label{app:popbalancemore}

The population balance equation is a phenomenological evolution equation for a probability density function in a predefined phase space describing a population of discrete entities. As such, the equation should respect the conservation laws governing this population. In \eqref{eqn:pbe_fbar_orig}, the total mass of gas in all the bubbles is conserved in the $\bs{x}$--$D$ phase space, except for buoyant degassing, the influx of gas due to processes like large-scale entrainment, and analogous sink terms for the gaseous mass in the limit of small bubble sizes. It has been implicitly assumed that a single parameter, $D$, suitably describes the size of the bubbles~\citep{Williams2}. As suggested in hypothesis~\ref{hyp:sphericity} in appendix~\ref{app:bubblehypo}, this is appropriate in a flow with a sufficiently high $\We_L$. Note that the population balance equation resembles the classical Liouville equation, except that no claim is made here about the divergence of the phase-space velocity field. One may also interpret \eqref{eqn:pbe_fbar_orig} as a generalized Boltzmann equation~\citep{Garrettson1,Carrica1,Solsvik1} where bubbles may split or be entrained in addition to colliding with one another. These various effects are subsumed in the generalized collision term, $H$.

Equation \eqref{eqn:pbe_fbar} describes the movement of gaseous mass in bubble-size space in small, localized regions of turbulent bubbly flows, albeit in a probabilistic manner. By the conservation of total mass of gas and statistical quasi-stationarity, $H(D)$ must satisfy
\begin{equation}
\int_0^\infty \diff D \: H(D) = R_d/\mathcal{V} + R_e/\mathcal{V} \underbrace{= 0}_{\substack{\text{statistical}\\\text{quasi-stationarity}}}
\label{eqn:pbe_masscons}
\end{equation}
in the limit of negligible buoyant degassing, where $R_d<0$ and $R_e>0$ are the volumetric rates of small-scale removal and large-scale addition, respectively. Noting that the left-hand side of \eqref{eqn:pbe_fbar} is the conservative form of the convective operator acting on $f(D)D^3$, \eqref{eqn:pbe_masscons} implies that the total amount of $f(D)D^3$ in the entire semi-infinite $D$-space cannot change except due to gas removal and/or addition, whose effects balance each other in the limit of statistical quasi-stationarity. Break-up and coalescence events do not generate or eliminate bubble mass, and thus do not contribute to the integral mass balance in \eqref{eqn:pbe_masscons}.

In \eqref{eqn:pbe_fbar2}, one may decompose $T_s(D) = T_d(D) + T_e(D) + T_g(D)$ into a small-scale sink kernel $T_d(D)$, a large-scale source kernel $T_e(D)$, and a sink kernel due to buoyant degassing $T_g(D)$, such that $\int_0^\infty \diff D \: T_d(D) = R_d/\mathcal{V} < 0$ and $\int_0^\infty \diff D \: T_e(D) = R_e/\mathcal{V} > 0$. If $T_d(D)$ and $T_e(D)$ are assumed to be active only at small and large $D$, respectively, and $T_c(D)$ and $T_g(D)$ are also assumed to be negligible, then an intermediate bubble-size subrange $L_\mathrm{H} \ll D \ll L$ emerges where $T_b(D) = 0$ as implied by hypothesis~\ref{hyp:intermediatesubrange} in appendix~\ref{app:bubblehypo}. Equations \eqref{eqn:pbe_locality} and \eqref{eqn:WbTb} further imply that $W_b$ is constant in this size subrange in the spirit of self-similarity.

\subsection{The model break-up kernel}\label{app:kernelmore}

Several properties of the probability distribution of child bubble sizes, $q_b(D_c|D_p)$, are introduced here~\citep{Ramkrishna1,MartinezBazan3}. The mechanics of break-up require~\citep{Valentas2}
\begin{equation}
q_b(D_c|D_p) = 0 \qquad \text{if } D_c \geq D_p,
\label{eqn:qb3}
\end{equation}
since a bubble cannot break to form bubbles larger than itself. Then, $q_b$ may be normalized such that
\begin{equation}
\int_0^{D_p} \diff D_c \: q_b(D_c|D_p) = \int_0^\infty \diff D_c \: q_b(D_c|D_p) = 2,
\label{eqn:qb1}
\end{equation}
where the factor of 2 arises from the assumption of binary break-up. As a result of this normalization, as well as the conservation of bubble mass, $q_b$ will also need to satisfy
\begin{equation}
\int_0^{D_p} \diff D_c \: q_b(D_c|D_p) D_c^3 = \int_0^\infty \diff D_c \: q_b(D_c|D_p) D_c^3 = D_p^3.
\label{eqn:qb2}
\end{equation}
Also, if a bubble of size $D_p$ breaks into two bubbles of sizes $D_1$ and $D_2$, then $q_b(D_1|D_p)/D_1^2 = q_b(D_2|D_p)/D_2^2$ by symmetry. This is more readily seen by observing equivalently that if a bubble of volume $D_p^3$ breaks into two bubbles of volumes $D_1^3$ and $D_2^3$, then $q_b\left(D_1^3|D_p^3\right) = q_b\left(D_2^3|D_p^3\right)$ by symmetry. An appropriate change in variables from $D^3$ to $D$ yields the desired relation. Using the properties of $q_b$ described above, one may verify that the model break-up kernel \eqref{eqn:Tb} satisfies \eqref{eqn:Tbcons} by direct substitution. One may also show via these properties of $q_b$ that $W_b$ satisfies
\begingroup
\allowdisplaybreaks
\begin{align}
W_b(D) &= \int_0^D \diff D_c \: T_b(D_c)\notag\\
&= \int_0^D \diff D_c \int_{D_c}^\infty \diff D_p \: q_b\left(D_c|D_p\right) g_b(D_p) f(D_p) D_c^3 - \int_0^D \diff D_c \: g_b(D_c) f(D_c) D_c^3\notag\\
&= \int_0^D \diff D_c \: D_c^3 \int_0^\infty \diff D_p \: q_b\left(D_c|D_p\right) g_b(D_p) f(D_p) - \int_0^D \diff D_p \: g_b(D_p) f(D_p) D_p^3\notag\\
&
\!\begin{multlined}[t]
= \int_0^D \diff D_c \: D_c^3 \int_0^\infty \diff D_p \: q_b\left(D_c|D_p\right) g_b(D_p) f(D_p) -{} \\
- \int_0^D \diff D_c \: D_c^3 \int_0^D \diff D_p \: q_b\left(D_c|D_p\right) g_b(D_p) f(D_p)
\end{multlined}\notag\\
&= \int_0^D \diff D_c \: D_c^3 \int_D^\infty \diff D_p \: q_b \left(D_c|D_p\right) g_b(D_p) f(D_p).
\end{align}
\endgroup


\section{Generalization of the bubble break-up formalism}\label{app:caveats}

\subsection{Non-binary break-up}\label{app:caveats-binary}

\begin{figure}
  \centerline{
\includegraphics[width=0.7\linewidth]{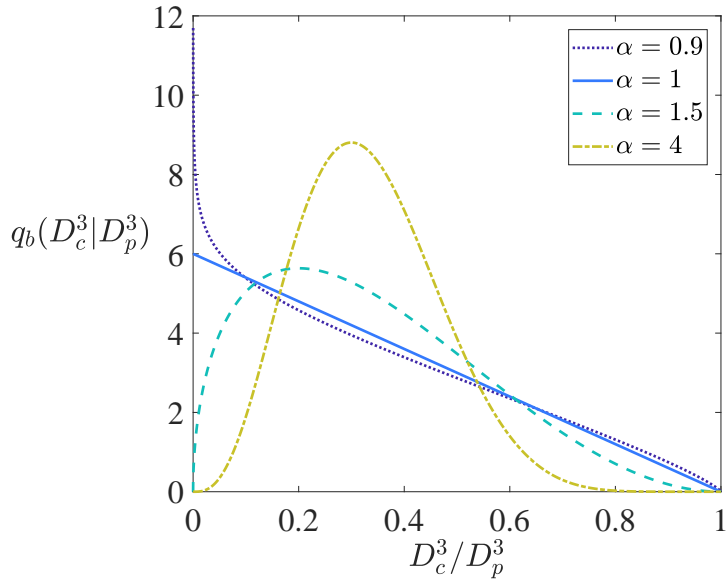}
}
  \caption{The generic beta distribution in $D^3$-space \eqref{eqn:betadis_gen} for various shape parameters $\alpha$. Here, $\beta$ is defined to be equal to $2\alpha$ so that $m=C=3$.} 
\label{fig:beta_gen}
\end{figure}

The constraints \eqref{eqn:qb1} and \eqref{eqn:qb2} need to be satisfied if bubbles in a system undergo only binary break-up events. These constraints need to be modified in the case of non-binary break-up. If the mean number of bubbles generated by a break-up event is $m \geq 2$, then the factor of 2 on the right-hand side of \eqref{eqn:qb1} will need to be replaced by $m$. The beta-distribution surrogate model in \S~\ref{sec:qbbeta} may be correspondingly modified to accommodate non-binary break-up. The generic beta distribution in bubble-volume space with two shape parameters $\alpha$ and $\beta$ takes the form
\begin{equation}
q_b\left(D_c^3|D_p^3\right) =
\begin{cases}
CD_c^{3(\alpha-1)}\left(D_p^3-D_c^3\right)^{\beta-1}D_p^{-3(\alpha+\beta-2)-3}/B(\alpha,\beta), &\qquad 0 \leq D_c^3 \leq D_p^3,\\
0, &\qquad D_p^3 < D_c^3.
\end{cases}
\label{eqn:betadis_gen}
\end{equation}
In order for the constraints \eqref{eqn:qb1}\textemdash with the right-hand side modified to $m$\textemdash and \eqref{eqn:qb2} to be satisfied, $\alpha$ and $\beta$ need to satisfy $(m-1)\alpha = \beta$, and $C$ needs to satisfy $C=m$. In the binary break-up limit $m=2$, one recovers $\alpha=\beta$ and $C=2$. The distribution \eqref{eqn:betadis_gen} is plotted in figure~\ref{fig:beta_gen} for several values of $\alpha$ in the case of $m=3$. Note that for the same $\alpha$, the large-$D_p$ and small-$D_c$ limits, \eqref{eqn:betaIpinf} and \eqref{eqn:betaIczero}, remain the same regardless of the value $m$ takes. This implies that the degrees of locality are comparable in two break-up processes that have different mean numbers of children bubbles but can be described with the same $\alpha$, which is the smaller of the two shape parameters characterizing the beta distribution. Also, the break-up process remains self-similar as long as both $\alpha$ and $\beta$ are constant over the size subrange of interest. These results imply that locality and self-similarity remain plausible in a break-up process that includes non-binary events.

\subsection{Non-self-similar break-up}\label{app:caveats-self-similar}

In the case of non-self-similar break-up, the scaling $g_b f \sim D_p^{-4}$ is no longer guaranteed to hold, as alluded to in \S~\ref{sec:spectralenergy}. However, the locality of the break-up flux appears to remain robust even in the absence of self-similarity. Consider the beta-distribution surrogate model in \S~\ref{sec:qbbeta} with the binary break-up assumption. Equation \eqref{eqn:betaIpinf} suggests that as long as the differential break-up rate $g_b f$ is a decreasing function of $D_p$ as $D_p \rightarrow \infty$, the break-up flux remains quasi-local for any permissible $\alpha$. In the worst-case scenario $\alpha \rightarrow 0$, $\gamma_p$ remains negative as long as the condition stated above holds true. More rigorously, the integral of $I_p$ with respect to $D_p$ from $D$ to $\infty$ is only defined if $I_p$ decays faster than $D_p^{-1}$. To ensure quasi-locality, one should then require that $g_b f$ also decays faster than $D_p^{-1}$. One may also argue the relative robustness of locality in the following manner: while a size-dependent $\alpha$ immediately results in a size-dependent $W_b$, \eqref{eqn:betaIpinf} and \eqref{eqn:betaIczero} suggest that the break-up flux may remain size local even if $\alpha$ is a function of the bubble size of interest. 

Recall from \S~\ref{sec:qbuniform} and \S~\ref{sec:qbbeta} that a self-similar $W_b$ is compatible with the statistical quasi-stationarity and quasi-homogeneity of the underlying system. Conversely, the absence of self-similarity suggests that the underlying system dynamics may not be statistically quasi-stationary. Consider \eqref{eqn:pbe_locality} in relation to the discussion in the preceding paragraph, which remarks that the break-up flux may remain size local even if it departs from self-similarity. This corresponds to the observation that the two terms of \eqref{eqn:pbe_locality} may still balance each other while being non-zero each. This, in turn, suggests that while an appropriate velocity $v_D$ may still be used to model a non-self-similar break-up process if there is sufficient locality, both $v_D$ and the bubble size distribution $f$ may become functions of time in the absence of self-similarity. In this case, a time-invariant power-law variation of $f$ cannot be assumed.


\section{The large-scale entrainment rate $Q$}\label{app:entrainment}

The gaseous volume entrainment rate per unit domain volume, $Q$, is typically assumed to be constant and imposed by integral-scale quantities like $u_L$ and $L$. It was observed in \S~\ref{sec:spectralenergy} that just as the large-scale energy production rate $\varepsilon$ is also assumed to be the turbulent kinetic energy cascade rate $\varepsilon$ in the turbulent energy cascade, the large-scale entrainment rate $Q$ and the bubble-mass cascade rate $W_b$ appear to be synonymous in the turbulent bubble break-up cascade. Two follow-up remarks are in order here. First, $Q$ and $\varepsilon$ are imposed by the large scales and may both depend on $u_L$ and $L$. Thus, $Q$ itself may appear to have an implicit dependence on $\varepsilon$, as remarked by~\citet{Deike1} and~\citet{Yu2}, who suggest that $Q$ is an increasing function of $\varepsilon$. More specifically, the quantity $\varepsilon$ in hypotheses \ref{hyp:bubbleuniversality} and \ref{hyp:intermediatesubrange} and figure~\ref{fig:cascade} may be equivalently replaced by $Q$ to no detriment. Second, recall from \S~\ref{sec:bubblecascade} that inertial effects dominate at large scales and capillary effects dominate at small scales. In a cascade mechanism with sufficient scale separation, large-scale quantities like $Q$ and $\varepsilon$ are unlikely to have implicit dependences on small-scale parameters like $\sigma/\rho_l$. Thus, in theories of bubble break-up that imply such a dependence in the sense that $f$ itself is proposed to be a function of $\sigma/\rho_l$, the underlying mechanism may not be self-similar due to the lack of scale separation. This is also a direct consequence of hypothesis~\ref{hyp:intermediatesubrange}: if an intermediate subrange of bubble sizes exists where the bubble dynamics are self-similar, then the corresponding bubble statistics are not a function of $\sigma/\rho_l$. It follows from appendix~\ref{app:caveats-self-similar} that a quasi-stationary power-law dependence should not be assumed in a system where the bubble dynamics at intermediate sizes depend on $\sigma/\rho_l$.


\bibliographystyle{jfm}
\bibliography{bubblecascade_R1}

\end{document}